\documentclass[extra,mreferee]{gji}
\usepackage{hyperref}
\usepackage{timet}
\usepackage{graphicx}
\usepackage{float}
\usepackage{amsmath,bm}
\usepackage{amssymb}
\usepackage{multirow}
\usepackage{lineno}
\usepackage{tikz}
\usepackage{mathtools}
\usepackage[textsize=tiny]{todonotes}
\usetikzlibrary{bayesnet}
\usepackage[final]{changes}
\definechangesauthor[color=red]{CG}

\title[Bayesian waveform-based calibration of high-pressure acoustic emission systems]
  {Bayesian waveform-based calibration of high-pressure acoustic emission systems with ball drop measurements}
\author[C.~Gu, U.~Mok, Y.~M.~Marzouk, G.~Prieto, F.~Sheibani, J.~B.~Evans, and B.~H.~Hager]
  {Chen Gu$^1$, Ulrich Mok$^1$, Youssef M. Marzouk$^{1,2}$, Germ\'an A Prieto Gomez$^3$, 
  \\\LARGE{\emph{Farrokh Sheibani$^1$, J. Brian Evans$^1$, and Bradford H. Hager$^1$}}\\
  $^1$ Earth Resources Laboratory, Department of Earth, Atmospheric, and Planetary Sciences, \\
 Massachusetts Institute of Technology, Cambridge, MA \emph{02139}, USA. Email: guchch@mit.edu\\
 $^2$ Aerospace Computational Design Laboratory, Department of Aeronautics and Astronautics\\
 Massachusetts Institute of Technology, Cambridge, MA \emph{02139}, USA.\\
 $^3$ Departamento de Geociencias, Universidad Nacional de Colombia, Bogot\'a, Colombia. \\
}

\begin{document}
\label{firstpage}
\maketitle

\begin{summary}
Acoustic emission (AE) is a widely used technology to study source mechanisms and material properties during high-pressure rock failure experiments. It is important to understand the physical quantities that acoustic emission sensors measure, as well as the response of these sensors as a function of frequency. This study calibrates the newly built AE system in the MIT Rock Physics Laboratory using a ball-bouncing system. Full waveforms of multi-bounce events due to ball drops are used to infer the transfer function of lead zirconate titanate (PZT) sensors in high pressure environments. Uncertainty in the sensor transfer functions is quantified using a waveform-based Bayesian approach. The quantification of \textit{in situ} sensor transfer functions makes it possible to apply full waveform analysis for acoustic emissions at high pressures.
\end{summary}


\section{Introduction}
The history of \emph{acoustic emission} (AE) dates back to the middle of the 20th century, before the term was coined in the work of \citet[]{schofield1963acoustic}. \citet[]{obert1942use} first detected subaudible noises emitted from rock under compression and attributed these signals to microfractures in the rock. \citet[]{kaiser1950study} recorded signals from the tensile specimens of metallic materials. Since the 1960s, much subsequent work has contributed to the development of AE techniques, which have been applied to diverse engineering and scientific problems \citep[]{drouillard1987introduction,drouillard1996history, grosse2008acoustic}.

AE is a useful tool to study the source mechanisms of ``labquakes'' and the three-dimensional structure of samples under diverse fracturing experimental conditions \citep[]{pettitt1998acoustic, schofield1963acoustic, ojala2004strain, Graham2010comparison, stanchits2011fracturing, w2013acoustic, fu2015experimental, hampton2015fracture, goodfellow2015hydraulic,li2017comparison, brantut2018time}. However, it is very difficult to use full waveforms of AE to infer AE source physics and sample structures, because AE amplitudes are affected by many factors (e.g., sensor coupling, frequency response of sensors, or incidence angle of ray paths) not related to the AE source or path effects. To determine the real physical meanings of the recorded AEs, careful calibration of their amplitudes is needed.

\citet{mclaskey2012acoustic} performed AE sensor calibration tests on a thick plate with two calibration sources (ball impact and glass capillary fracture) to estimate instrument response functions. \citet{ono2016calibration} demonstrated detailed sensor calibration methods, including face-to-face, laser interferometry, Hill-Adams equation, and tri-transducer methods. \citet{yoshimitsu2016geometric} combined laser interferometry observations and a finite difference modeling method to characterize full waveforms from a circular-shaped transducer source through a cylindrical sample. However, these calibration methods only work under ambient conditions, and not within a pressure vessel where rock physics experiments are sometimes carried out. To calibrate the AE amplitudes under high-pressure conditions, \citet{kwiatek2014seismic} proposed an \textit{in situ} ultrasonic transmission calibration (UTC) method to correct relative amplitudes under high pressure. \citet[]{mclaskey2015robust} developed a technique to calibrate a high-pressure AE system using \textit{in situ} ball impact as a reference source. This design enabled the determination of absolute source parameters with an \textit{in situ} accelerometer.                                         

This study aims to advance these calibration methodologies by quantifying the uncertainty of sensor transfer functions using a waveform-based Bayesian approach. Instead of using the waveform of a single ball bounce, our approach is able to use the waveforms of multi-bounce events. Inferring an \textit{in situ} sensor transfer function, and its associated uncertainty, makes it possible to apply full waveform analysis for acoustic emissions under high-pressure conditions. The method is tested using the newly built AE system of the MIT Rock Physics Laboratory.

\section{Methodology}
\subsection{Experimental Setup and AE Data}
\label{sec:exp}
\begin{figure}
\begin{center}
\includegraphics[scale=0.65]{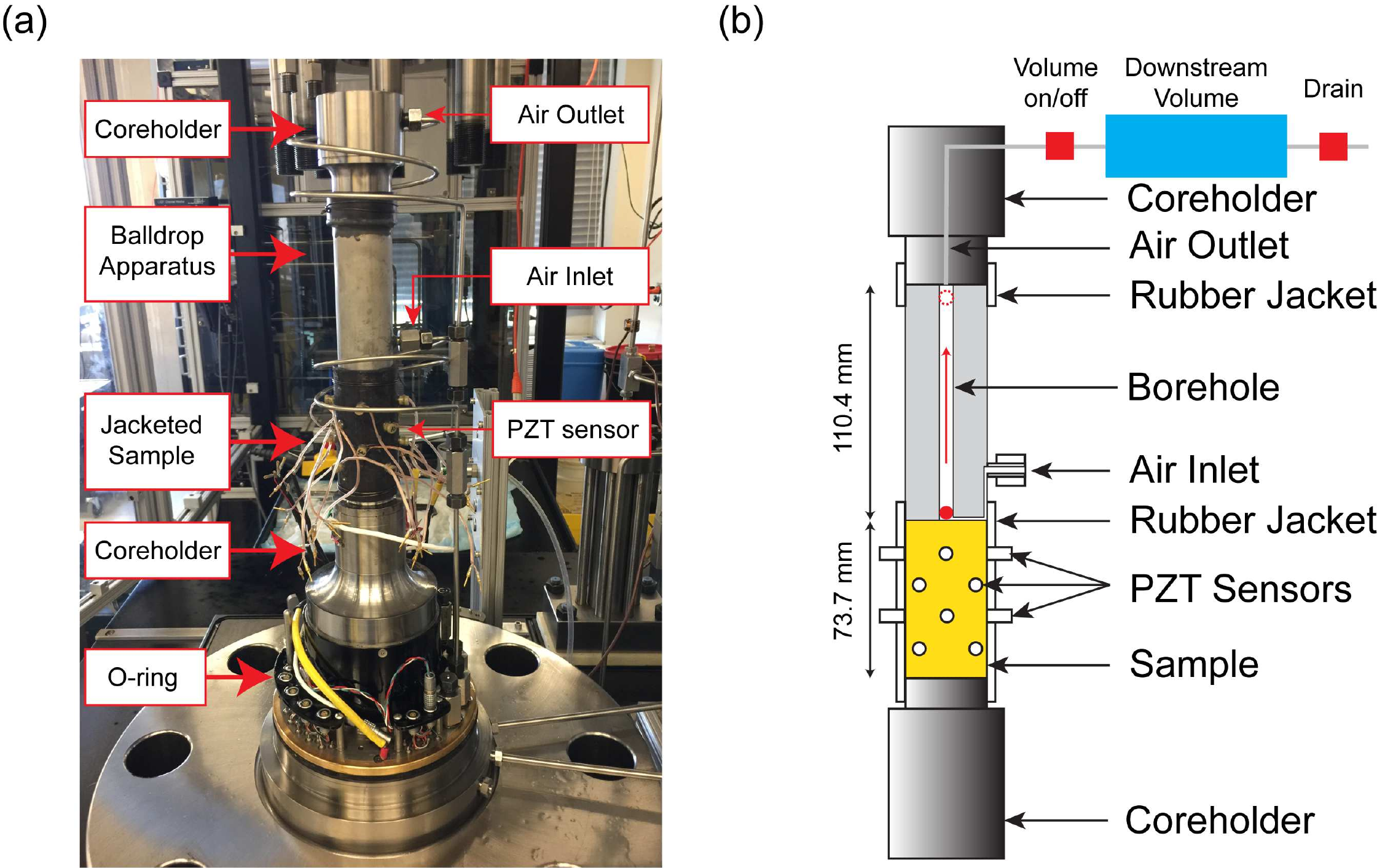}
\centering
\caption{(a) Photo of sample assembly before closing the pressure vessel. (b) Schematic of cross-section of the ball drop apparatus and the instrumented sample.}
\label{fig:exp_setup}
\end{center}
\end{figure}

\begin{figure}
\begin{center}
\includegraphics[scale=0.75]{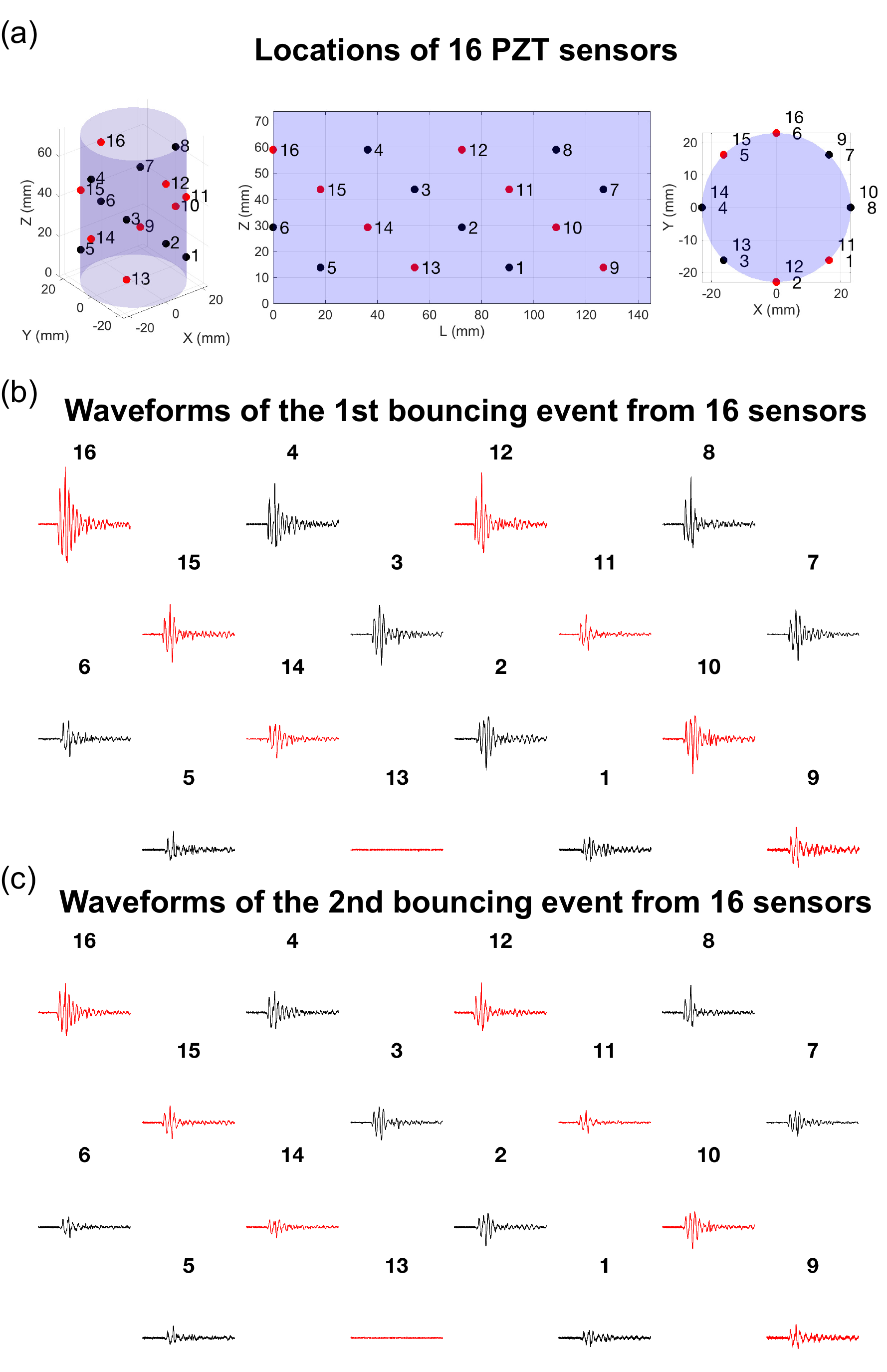}
\centering
\caption{(a) Locations of 16 PZT sensors. (b) Example waveforms from 16 sensors for the 1st ball bounce \added[id=CG]{at cp = 30 MPa and ds = 10 MPa}.(b) Example waveforms from 16 sensors for the 2nd ball bounce \added[id=CG]{at cp = 30 MPa and ds = 10 MPa}. Black and red denote sensors and corresponding received signals on two different boards.}
\label{fig:sensor_distr}
\end{center}
\end{figure}

The ball drop apparatus to conduct the \textit{in situ} ball drop experiment is shown in Figure~\ref{fig:exp_setup}. A steel ball (radius $R_1=3.18$ mm) placed in a tube is lifted to the top by air blown into the tube. After the air is cut off, the ball drops and hits the surface of a titanium cylinder (marked as ``sample'' in Figure~\ref{fig:exp_setup}), bouncing a few times. The diameter of the titanium cylinder is 46.1 mm and the length is 73.7 mm. Sixteen lead zirconate titanate (PZT) sensors are attached to the surface of the titanium cylinder (Figure \ref{fig:sensor_distr}). We stack a nonpolarized PZT piezoceramic disk, a polarized PZT piezoceramic disk, and a titanium disk adapter together to make one sensor. The diameters of the polarized and nonpolarized PZT piezoceramic disks are 5.00 mm and the thicknesses are 5.08 mm. The resonance frequency is 1 MHz. The titanium disk adapter has a diameter of 5.00 mm and a thickness of 4.00 mm. The side of the titanium disk adapter contacting the polarized PZT piezoceramic disk is machined to be flat, and the other side contacting the cylindrical sample is machined to be concave, to better fit the curved cylindrical side surface. \replaced[id=CG]{We increase the confining pressures (cp) and differential stresses (ds) gradually to improve the coupling between the PZT sensors and the sample. The \textit{in situ} ball drop experiments are conducted at varied cp and ds. High-quality AE data are observed at: (1) cp = 10 MPa, ds = 6 MPa; (2) cp = 20 MPa, ds = 10 MPa; (3) cp = 30 MPa, ds = 10 MPa. Then we decrease both cp an ds to ambient conditions and conduct one more ball drop experiment as the baseline measurements.}{The ball drop experiment is conducted at a confining pressure of 30 MPa and a differential pressure of 10 MPa.}

\begin{figure}
\begin{center}
\includegraphics[scale=0.7]{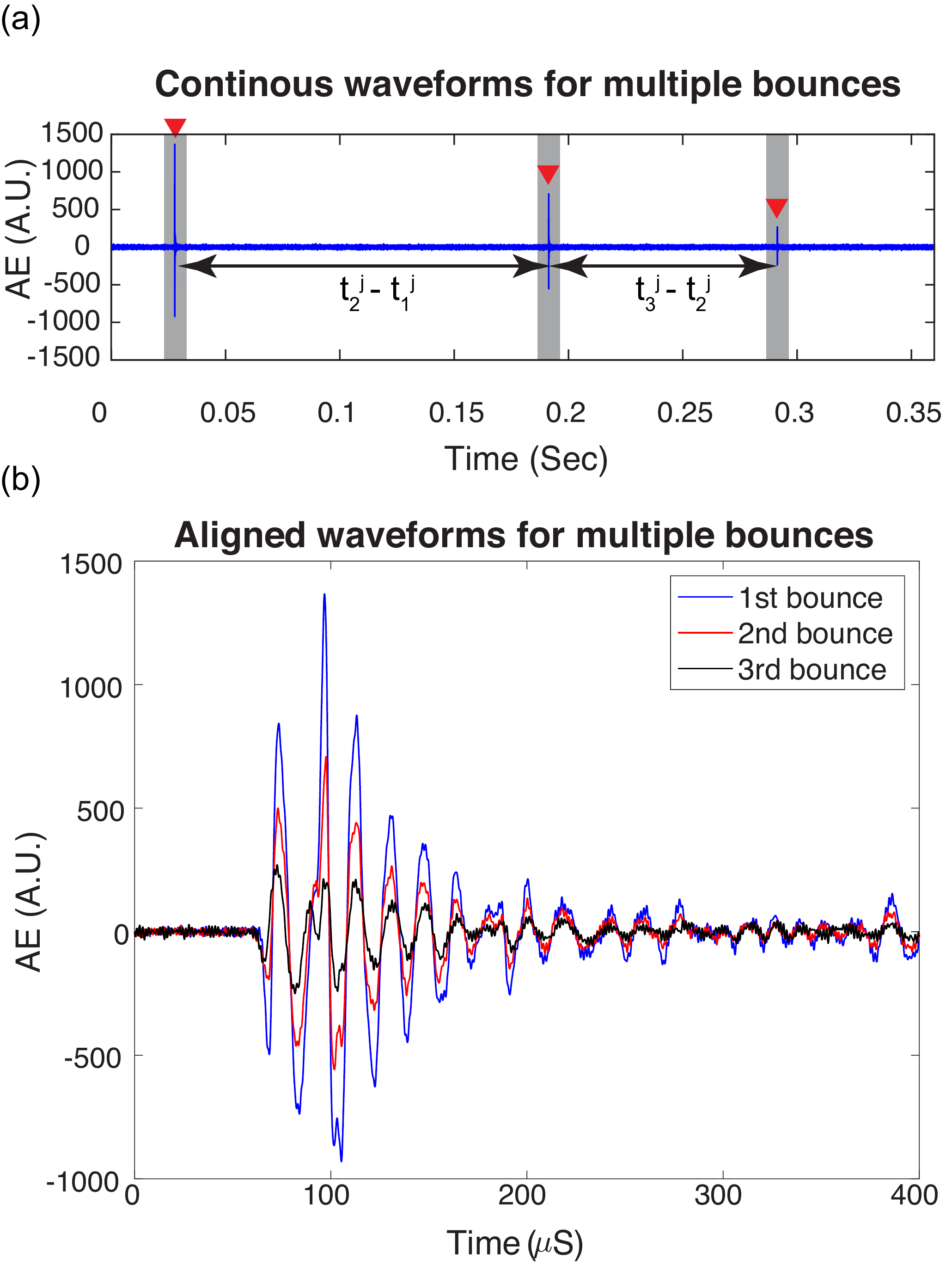}
\centering
\caption{(a) Continuous waveforms containing the first three bouncing events of sensor 16 \added[id=CG]{at cp = 30 MPa and ds = 10 MPa}. The grey shadow areas denote time windows of bouncing events used for cross-correlation. (b) Aligned waveforms of three continuous bouncing events of sensor 16 \added[id=CG]{at cp = 30 MPa and ds = 10 MPa}. }
\label{fig:bouncing_wfm}
\end{center}
\end{figure}

The AE data are continuously recorded and streamed to a hard drive at a sampling rate of 12.5 MHz, preprocessed by the STA/LTA algorithm to detect events due to ball bounces \citep[]{swindell1977station, mcevilly1982asp, earle1994characterization}. The truncated waveforms of the first and second bounces from 16 sensors due to one ball drop experiment at cp = 30 MPa and ds = 10 MPa are shown in Figure \ref{fig:sensor_distr}. We implement the Akaike information criterion (AIC) algorithm to automatically pick the P arrival time $t_1^j$ for the truncated waveforms of the first bouncing event \citep[]{maeda1985method, kurz2005strategies}. Then we align the waveforms from the later bounces and the first bounce by cross-correlation. An example of continuous waveforms containing the first three bouncing events of sensor 16 \added[id=CG]{at cp = 30 MPa and ds = 10 MPa} is shown in Figure~\ref{fig:bouncing_wfm}(a). The aligned waveforms of three bouncing events are shown in Figure~\ref{fig:bouncing_wfm}(b). The absolute P arrival time $t_k^j$ of the $k$th bouncing event at sensor $j$ can be calculated by adding the time lag between the waveforms of the first and the $k$th bouncing events to $t_1^j$. The time intervals between all the bounces recorded by sensor $j$ can thus be collected as
\begin{equation}
    \bm{\delta t}^j = [t_2^j - t_1^j, t_3^j - t_2^j, \ldots, t_{k+1}^j - t_k^j, \ldots, t_n^j - t_{n-1}^j],
    \label{eqn:dt_j}
\end{equation}
where $n$ is the total number of bounces.

\begin{figure}
\begin{center}
\includegraphics[scale=0.7]{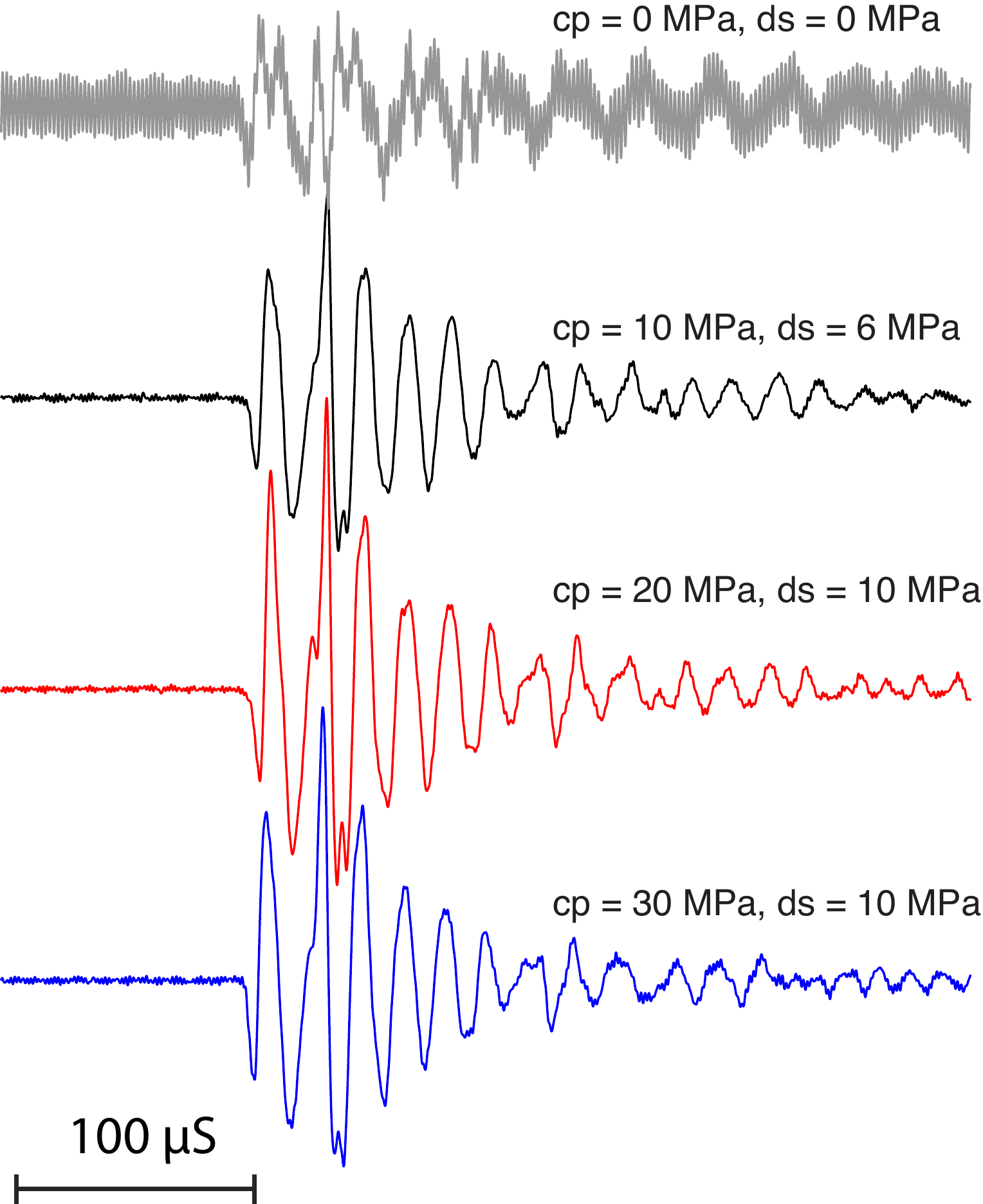}
\centering
\caption{\added[id=CG]{Waveforms at sensor 16 due to the first bounce of the ball drop at: (1) cp = 10 MPa, ds = 6 MPa; (2) cp = 20 MPa, ds = 10 MPa; (3) cp = 30 MPa, ds = 10 MPa; and (4) cp = 0 MPa, ds = 0 MPa. The waveforms are normalized by the maximum amplitues of all waveforms.}}
\label{fig:wfm_compare_cp_ds}
\end{center}
\end{figure}

\added[id=CG]{The same preprocessing method has also been applied to the AE data at other cp and ds. Under each conditions, we compared the waveforms at sensor 16 due to the first bounce of the ball drop at: (1) cp = 10 MPa, ds = 6 MPa; (2) cp = 20 MPa, ds = 10 MPa; (3) cp = 30 MPa, ds = 10 MPa; and (4) cp = 0 MPa, ds = 0 MPa in Figure \ref{fig:wfm_compare_cp_ds}. Because higher confining pressures improve the coupling between sensors and the sample, resulting in smaller noise and larger amplitude response of sensors, the waveforms at high pressures show a smaller noise lever compared to those at ambient conditions. For the same reason, the waveforms at cp = 20 MPa, ds = 10 MPa, and cp = 30 MPa,  ds = 10 MPa have larger amplitudes than those at cp = 10 MPa, ds = 6 MPa. However, when the confining pressure increases beyond a critical level, higher confining pressures do not affect the noise level and sensor response, as is illustrated in the almost identical waveforms at cp = 20 MPa, ds = 10 MPa, and cp = 30 MPa, ds = 10 MPa.}

\subsection{Bouncing time and waveform modeling}
\label{sec:forward}
To model the time interval between bounces, we first assume that after each bounce, the rebound velocity decreases to a fraction $a$ (the rebound coefficient) of the incident velocity; then the velocity after the $k$th bounce is
\begin{equation}
    v_k = a^k v_0.
\end{equation}
The time interval between the $(k+1)$th and the $k$th bounce is then

\begin{equation}
    \tilde{t}_{k+1}^j - \tilde{t}_k^j = \frac{2v_k}{g} = \frac{2a^k v_0}{g},
\end{equation}
where $g$ is the acceleration of gravity.
The theoretical bouncing time intervals $\bm{\delta t}^j = [\tilde{t}^j_2 - \tilde{t}^j_1, \ldots, \tilde{t}^j_n - \tilde{t}^j_{n-1}]$ can then be modeled as
\begin{equation}
    \bm{\delta t_m}^j = \left[\frac{2av_0}{g}, \frac{2a^2v_0}{g}, \ldots, \frac{2a^{n-1} v_0}{g}\right]
    \label{eq:dt}
\end{equation}
Now, modeling the mismatch between the modeled and measured time intervals with additive noise $\bm{e_t}^j$, the bouncing time interval data $\bm{\delta t}^j$ is represented as
\begin{equation}
   \bm{\delta t}^j = \bm{\delta t_m}^j  + \bm{e_t}^j.
   \label{eqn:bouncingdata}
\end{equation}

The waveform recorded at receiver $j$ due to the $k$th ball bounce, $\bm{o}^{j,k}(t)$, can be written as
\begin{equation}
\bm{o}^{j,k}(t) = L[I[\bm{u}^{j,k}(\bm{r}^j, 
t)]] = L[I[f_k(t)*\bm{G}(\bm{r}^j,t)]],
\label{eqn:convolution}
\end{equation}
where $\bm{u}^{j,k}(\bm{r}^j, t)$ is the input displacement at receiver $j$ due to the $k$th bounce of the ball, $f_k(t)$ is the loading function of the $k$th bounce of the ball, $\bm{G}(\bm{r}^j,t)$ is the Green's function representing the impulse response of the sample at receiver $j$, $t$ is the time, $\bm{r}^j$ is the vector directed from the bouncing ball source to receiver $j$, $I$ is the incident angle correction, and $L$ is a linear operator, assuming that the response function of the PZT transducer can be modeled as a linear time-invariant (LTI) system.
Based on previous studies of ball collisions \citep{mclaskey2012acoustic,mclaskey2015robust}, the loading function can be represented as
\begin{equation}
    \begin{split}
        &f_k(t) = -F_{\text{max},k}\sin\left(\frac{\pi t}{t_c}\right)^{3/2}, \quad 0\leq |t| \leq t_c,\\
        &f_k(t) = 0, \quad \text{otherwise},   
    \end{split}
\end{equation}
where $F_{\text{max},k}$ is the maximum loading force of the $k$th ball bounce and $t_c$ is the total loading time, which is the entire contact time between the ball and the top surface of the sample. The $F_{\text{max},k}$ and $t_c$ are modeled as
\begin{eqnarray}
&F_{\text{max},k}&=1.917\rho_1^{3/5}(\delta_1 + \delta_2)^{-2/5} R_1^2 v_{k-1}^{6/5},\\ 
&\delta_q &= \frac{1-\mu_q^2}{\pi E_q}, \quad q = 1,2\\
&t_c&=1/f_c=4.53(4\rho_1\pi(\delta_1 + \delta_2)/3)^{2/5}R_1v_{k-1}^{-1/5},
\end{eqnarray}
where $\rho_q$, $E_q$, $\mu_q$ are the density, Young's modulus, and Poisson's ratio of the $q$th material, respectively ($q=1$ refers to the steel ball and $q=2$ refers to the titanium sample). In this experiment, $\rho_1=8050~\text{kg}/\text{m}^3$, $E_1=180.0~\text{GPa}$, $\mu_1=0.305$, $\rho_2=4506~\text{kg}/\text{m}^3$, $E_2=113.8~\text{GPa}$, and $\mu_2=0.32$. $v_{k-1}$ is the incident velocity of the $k$th bounce of the ball.


$G_{i3}(\bm{r}^j,t)$ is the $i$th ($i = 1, 2, 3$, corresponding to three axes) component of displacement at a generic $(\bm{r}^j,t)$, for an impulsive point force source in the $x_3$ direction, i.e., the vertical direction. The $i$th component of displacement due to the $k$th bounce, $u_i^{j,k}(\bm{r}^j, t)$, is represented as \citep{aki2002quantitative}
\begin{equation}
\label{eq:3C_disp}
\begin{split}
u_i^{j,k}(\bm{r}^j, t) = & f_k(t) * G_{i3}(\bm{r}^j,t), \\
               = &\frac{1}{4\pi \rho_2} (3\gamma_i^j \gamma_3^j - \delta_{i3})\frac{1}{(r^j)^3} \int_{{r^j}/{V_P}}^{{r^j}/{V_S}}\tau f_k(t-\tau)d\tau \\
                 &+\frac{1}{4\pi \rho_2 V_P^2} \gamma_i^j \gamma_3^j \frac{1}{r^j} f_k\left(t - \frac{r^j}{V_P}\right) - \frac{1}{4 \pi \rho_2 V_S^2}(\gamma_i^j \gamma_3^j - \delta_{i3})\frac{1}{r^j}f_k\left(t - \frac{r^j}{V_S}\right),
\end{split}
\end{equation}
where $V_P=6011.6~\text{m}/\text{s}$ and $V_S=3093.0~\text{m}/\text{s}$ are the P wave velocity and S wave velocity of the titanium sample, respectively; $r^j$ is the norm of the vector from the source to sensor $j$; $\gamma_i^j$ is the directional cosine between this vector and the $i$th coordinate axis; and $\delta_{i3}$ is the Kronecker delta. \added[id=CG]{This Green's function is for a homogeneous, isotropic, unbounded medium. We use this approximation is because we do not observe coherent signals due to possible reflections from boundaries in the data. Second, the finite difference modeling in Appendix \ref{sec:bd_wave} shows that for our calibration system with a Titanium sample, this homogeneous, isotropic, unbounded medium approximation produces almost identical waveforms compared to the system with a Titanium-Steel boundary on the top. The ball drop apparatus is made of Steel.}

The incidence angle dependence of the sensor is assumed to be a cosine function, i.e., 
\begin{equation}
    I[\bm{u}^{j,k}(\bm{r}^j, t)]=u_{\perp}^{j,k}(\bm{r}^j, t)=\sum_{i=1}^3u_i^{j,k}(\bm{r}^j, t)\xi_i^j,
    \label{eqn:directional}
\end{equation}
where $\xi_i^j$ is the directional cosine of the normal vector of sensor $j$, i.e., $[r_1^j, r_2^j, 0]$. \added[id=CG]{This cosine approximation is justified in Appendix \ref{sec:amp_inc}.}

The frequency response function of sensor $j$ is modeled by
\begin{equation}
    \mathfrak{R}^j(\omega) = \frac{-C\omega^2}{\omega^2 + 2i\varepsilon^j\omega - (\omega_s^j)^2},
    \label{eq:resp}
\end{equation}
where $\omega_s^j$ is the resonance frequency, $\varepsilon^j$ is the damping coefficient of sensor $j$, and $C$ is the conversion constant with units $\text{count}/\text{m}$. \added[id=CG]{This simple frequency response function of a damped oscillator can fully describe the resonance and damping effects of PZT sensors according to the full waveform matching, so we do not include high-fidelity PZT sensor modeling method in our model \citep{baek2010modeling}. Equation~\eqref{eq:resp} has also been used to model the frequency-response of an inertial seismometer \citep{aki2002quantitative}.}

Then the noise-free signal at sensor $j$ due to the $k$th bounce can be represented as
\begin{equation}
    O^{j,k}(\omega) = \mathfrak{R}^j(\omega)U_{\perp}^{j,k}(\omega)
    \label{eqn:fouriersensoroutput}
\end{equation}
in the frequency domain, and 
\begin{equation}
    o^{j,k}(t) = \mathfrak{r}^j(t)*u_{\perp}^{j,k}(t)
    \label{eqn:timeseriesoutput}
\end{equation}
in the time domain, where $\ast$ represents the convolution operator. In \eqref{eqn:fouriersensoroutput}, $U_{\perp}^{j,k}(\omega)$ is simply the Fourier transform of $u_{\perp}^{j,k}(t)$ \eqref{eqn:directional}. Similarly, $\mathfrak{r}^j(t)$ is the inverse Fourier transform of $\mathfrak{R}^j(\omega)$ \eqref{eq:resp}. 
Concatenating waveforms from all the bounces, along with their corresponding noise perturbations $e^{j,k}(t)$, the data at receiver $j$ can be modeled as
\begin{equation}
\left[
\begin{array}{c}
     d^{j,1}(t)\\
     d^{j,2}(t)\\
     \vdots\\
     d^{j,k}(t)\\
     \vdots\\
     d^{j,n}(t)\\
\end{array}
\right]
     = 
\left[
\begin{array}{c}
     o^{j,1}(t)\\
     o^{j,2}(t)\\
     \vdots\\
     o^{j,k}(t)\\
     \vdots\\
     o^{j,n}(t)\\
\end{array}
\right]+
\left[
\begin{array}{c}
     e^{j,1}(t)\\
     e^{j,2}(t)\\
     \vdots\\
     e^{j,k}(t)\\
     \vdots\\
     e^{j,n}(t)\\
\end{array}
\right],
\end{equation}
which can be written more compactly as
\begin{equation}
    d^j(t)=o^j(t)+e^j(t),
    \label{eq:wavedata}
\end{equation}
and, after time discretization, as
\begin{equation}
    \bm{d}^j=\bm{o}^j+\bm{e}^j.
    \label{eq:wavedatadiscrete}
\end{equation}

\subsection{Bayesian formulation and posterior sampling}
\label{sec:bayes}
In principle, one could use a Bayesian hierarchical model to represent the entire ball drop system, given bouncing time interval data $\bm{\delta t} \coloneqq \{ \bm{\delta t}^j \}_{j=1}^{16}$ and waveform data $\{\bm{d}^j\}_{j=1}^{16}$ from all 16 sensors. The resulting posterior density is: 
\begin{equation}
\begin{split}
    &P\left(\{v_0^j, a^j, \omega_s^j, \varepsilon^j\}_{j=1}^{16}, v_0, a, \sigma_t^2|\{\bm{d}^j\}_{j=1}^{16}, \bm{\delta t} \right)\\
    \propto & \ P(\bm{\delta t} | v_0, a, \sigma_t^2)P(v_0)P(a)P(\sigma_t^2)\\
    &\left ( \prod\limits_{j=1}^{16}P(\bm{d}^j|v_0^j, a^j, \omega_s^j, \varepsilon^j)\right) 
    \left ( \prod\limits_{j=1}^{16}P(v_0^j, a^j|v_0, a) \right )
    \left ( \prod\limits_{j=1}^{16}P(v_0^j)P(a^j)P(\omega_s^j)P(\varepsilon^j) \right ),
\end{split}
\end{equation}
We explain this model, and the terms above, as follows. First, there is in principle a single true value of the ball's initial incident velocity and rebound coefficient, represented by the ``master'' parameters $v_0$ and $a$. All of the bouncing time intervals $\bm{\delta t}$ should depend on these values; this relationship is encoded in the conditional probability density $P(\bm{\delta t} | v_0, a, \sigma_t^2)$. Here $\sigma_t^2$ is the variance of the noise $\bm{e}_t^j$ in \eqref{eqn:bouncingdata}, which we also wish to infer. 
The full waveforms $\bm{d}^j$ at each receiver $j$ also depend on the ball velocity and rebound coefficient, however, as these are needed to determine the loading function for each individual bounce $k=1, \ldots, n$. Due to noise and unmodeled dynamics, these waveforms may be better represented by slightly different local bouncing parameters, $v_0^j$ and $a^j$, at each receiver $j$. To relate the local bouncing parameters $v_0^j$ and $a^j$ with the master parameters $v_0$ and $a$, as is typical in Bayesian hierarchical modeling \citep{gelman2013bayesian}, the model above uses the conditional distributions $P(v_0^j, a^j|v_0, a)$. The relationship between the local bouncing parameters and the full waveforms is encoded in the likelihood $P(\bm{d}^j|v_0^j, a^j, \omega_s^j, \varepsilon^j)$. 

To simplify and decouple this inference problem, however, we can ignore the relationship between the master $(v_0, a)$ and $(v_0^j, a^j)$, i.e., we can assume $P(v_0^j, a^j|v_0, a) \approx P(v_0^j, a^j)$. Then the master parameters become irrelevant and we can infer parameters $\bm{X}=[v_0^j, a^j, \omega_s^j, \varepsilon^j]$ and a variance $(\sigma_t^j)^2$ for each sensor separately. This assumption is reasonable because there is a considerable amount of data/information at each sensor; thus, there is little to be gained by ``sharing strength'' via the common parameters $(v_0, a)$.  
For sensor $j$, the posterior probability density $P(v_0^j, a^j, \omega_s^j, \varepsilon^j,(\sigma_t^j)^2|\bm{d}^j, \bm{\delta t})$ is then written as
\begin{equation}
\begin{split}
    P(v_0^j, a^j, \omega_s^j, \varepsilon^j, (\sigma_t^j)^2|\bm{d}^j, \bm{\delta t}) \propto \  &P(\bm{\delta t}|v_0^j, a^j,(\sigma_t^j)^2) P(\bm{d}^j|v_0^j, a^j, \omega_s^j, \varepsilon^j) \\
     &P(v_0^j)P(a^j)P(\omega_s^j)P(\varepsilon^j) P((\sigma_t^j)^2).
\label{eqn:simplifiedposterior}
\end{split}
\end{equation}
Figure \ref{fig:MCMC_procedure} represents the hierarchical Bayesian model and the simplified Bayesian model graphically, as Bayesian networks.

Now we define the specific prior and likelihood terms in the  posterior probability density function \eqref{eqn:simplifiedposterior}. We use uniform prior  distributions for $v_0^j$, $a^j$, $\omega_s^j$, and $\varepsilon^j$, i.e.,
\begin{equation}
\begin{split}
    &v_0^j \sim \mathcal{U}(1.0, 1.5)\ \text{m/s}, \quad a^j \sim \mathcal{U}(0.5, 0.9),\\ 
    &\omega_s^j \sim \mathcal{U}(100, 500) \ \text{kHz},\quad \varepsilon^j \sim \mathcal{U}(10, 50)\ \text{kHz},
\end{split}
\end{equation}
and a normal distribution for $(\sigma_t^j)^2$,
\begin{equation}
    (\sigma_t^j)^2 \sim \mathcal{N}(10^{-10}, 10^{-22})\ \text{s}^2.
    \label{eq:hyperprior}
\end{equation}

The likelihood functions $P(\bm{\delta t}|v_0^j, a^j)$ and $P(\bm{d}^j|v_0^j, a^j, \omega_s^j, \varepsilon^j)$ depend on the probability distributions of $\bm{e_t} = \bm{\delta t} - \bm{\delta t_m}$ \eqref{eqn:bouncingdata} and $\bm{e}^j = \bm{d}^j - \bm{o}^j$ \eqref{eq:wavedata}, respectively. In this paper, we assume that both errors are Gaussian, with zero mean and diagonal covariance matrices $\bm{\Sigma_t}$ and $\bm{\Sigma}^j$, respectively. The diagonal entries of $\bm{\Sigma_t}$ are
\begin{equation}
    \Sigma_{t,ii}=(\sigma_t^j)^2, \quad i=1,2,\ldots, N_{\delta t},
\end{equation}
where $N_{\delta t}$ is the total number of bouncing time intervals, collected over all the sensors.
The diagonal entries of $\bm{\Sigma}^j$ are
\begin{equation}
    \Sigma^j_{ii}=(\sigma^j)^2, \quad i=1,2,\ldots, N
\end{equation}
where $N$ is the total number of data samples (time discretization points) for the waveforms $\bm{d}^j$. Then the likelihood functions can be written as
\begin{eqnarray}
    P(\bm{\delta t}|v_0^j, a^j,(\sigma_t^j)^2) = \frac{1}{\sqrt{(2\pi)^{N_{\delta t}} \det \bm{\Sigma_t}}} \exp{\left[-\frac{1}{2}(\bm{\delta t} - \bm{\delta t_m})^T \bm{\Sigma_t}^{-1}(\bm{\delta t} - \bm{\delta t_m})\right]},\\
    P(\bm{d}^j|v_0^j, a^j, \omega_s^j, \varepsilon^j) = \frac{1}{\sqrt{(2\pi)^N \det \bm{\Sigma}^j}} \exp{\left[-\frac{1}{2}(\bm{d}^j - \bm{o}^j)^T (\bm{\Sigma}^j)^{-1}(\bm{d}^j - \bm{o}^j)\right]}.
\end{eqnarray}

\begin{figure}
(a)
\begin{center}
\begin{tikzpicture}
  \node[obs]                               (dt) {$\bm{\delta t}$}; 
  \node[latent, right=2cm of dt] (v0) {$v_0$}; 
  \node[latent, right=2of dt, yshift=-1.2cm]  (a) {$a$}; 
  \node[latent, left=2cm of dt]            (et) {$\bm{e}_t$}; 
  \node[latent, left=2cm of dt, yshift=-1.2cm] (st) {$\sigma_t$}; 
  \node[obs, right=7cm of dt, yshift=-1.8cm]                               (d) {$\bm{d}^j$};
  \node[latent, right=4cm of dt] (v0j) {$v_0^j$};
  \node[latent, right=4cm of dt, yshift=-1.2cm]  (aj) {$a^j$};
  \node[latent, right=4cm of dt, yshift=-2.4cm] (omegasj) {$\omega_s^j$};
  \node[latent, right=4cm of dt, yshift=-3.6cm]  (varepsilonj) {$\varepsilon^j$};
  \node[latent, right=2cm of d]            (e) {$\bm{e}^j$}; 
  \edge {v0,a,et,st} {dt} ; %
  \edge {v0j,aj,e,omegasj,varepsilonj} {d} ;
  \edge {v0} {v0j};
  \edge {a} {aj};
  \plate {} {(d)(v0j)(aj)(omegasj)(varepsilonj)(e)} {$j=1\ldots16$} ;
\end{tikzpicture}
\end{center}
(b)
\begin{center}
\begin{tikzpicture}
  \node[obs]                               (dt) {$\bm{\delta t}$}; 
  \node[latent, left=2cm of dt]            (etj) {$\bm{e}_t^j$}; 
  \node[latent, left=2cm of dt, yshift=-1.2cm] (stj) {$\sigma_t^j$}; 
  \node[obs, right=7cm of dt, yshift=-1.8cm]                               (d) {$\bm{d}^j$};
  \node[latent, right=4cm of dt] (v0j) {$v_0^j$};
  \node[latent, right=4cm of dt, yshift=-1.2cm]  (aj) {$a^j$};
  \node[latent, right=4cm of dt, yshift=-2.4cm] (omegasj) {$\omega_s^j$};
  \node[latent, right=4cm of dt, yshift=-3.6cm]  (varepsilonj) {$\varepsilon^j$};
  \node[latent, right=2cm of d]            (e) {$\bm{e}^j$}; 
  \edge {v0j,aj,etj,stj} {dt} ; %
  \edge {v0j,aj,e,omegasj,varepsilonj} {d} ;
  \plate {} {(d)(v0j)(aj)(omegasj)(varepsilonj)(e)(dt)(stj)} {$j=1\ldots16$} ;
\end{tikzpicture}
\end{center}
\caption{Schematic of MCMC procedure. (a) The hierarchical model; (b) the decoupled model.}
\label{fig:MCMC_procedure}
\end{figure}

The variance parameter $(\sigma_t^j)^2$ represents a tradeoff between the influence of the bouncing time interval data and the waveform data. As encoded in the posterior distribution \eqref{eqn:simplifiedposterior}, it is inferred from both sets of data, for each sensor $j$. The variance parameter $(\sigma^j)^2$, on the other hand, is not inferred within the Bayesian model, but is determined based on the noise level of the waveform data. In particular, we set it to be a fraction $\alpha^j \in(0, 1)$ of the power of the waveform data $\bm{d}^j$, i.e., 

\begin{equation}
    (\sigma^j)^2 = \alpha^j\frac{1}{T}\int\limits_0^T (d^j(t))^2 \, dt \approx \alpha^j\frac{1}{T}\sum\limits_{i=1}^N \Delta t (d_i^j)^2 = \alpha^j\frac{1}{N}\sum\limits_{i=1}^N (d_i^j)^2,
    \label{eq:sigmaj}
\end{equation}
where $T$ is the total time length of waveform data, $N = T / \Delta t$ is the number of time discretization points, and $d^j_i = d^j(t_i)$. We estimate $\alpha^j$, the ratio of noise and waveform data power for sensor $j$, as
\begin{equation}
    \alpha^j = \frac{\frac{1}{T_b}\int\limits_0^{T_b}b^2(t)dt}{\frac{1}{T}\int\limits_0^Td^2(t)dt} \approx \frac{\frac{1}{T_b}\sum\limits_{i=1}^{N_b} \Delta t (b_i^j)^2}{\frac{1}{T}\sum\limits_{i=1}^N \Delta t (d_i^j)^2} = \frac{N\sum\limits_{i=1}^{N_b} (b_i^j)^2}{N_b\sum\limits_{i=1}^{N_b} (d_i^j)^2},
    \label{eq:alphaj}
\end{equation}
where $b(t)$ is the recorded noise before the first P arrival of the first bouncing event, $T_b$ is the time length of the noise window, and $N_b = T_b / \Delta t$ is the number of noise samples. Substituting \eqref{eq:alphaj} into \eqref{eq:sigmaj}, we obtain
\begin{equation}
    \left(\sigma^j\right)^2 = \frac{1}{N_b} \sum\limits_{i=1}^{N_b} (b_i^j)^2.
    \label{eq:sigma2}
\end{equation}

For each sensor $j$, we use Markov chain Monte Carlo (MCMC) sampling to characterize the posterior distribution given by \eqref{eqn:simplifiedposterior}. We use an independence proposal from the prior to update $(\sigma_t^j)^2$ and a 
Gaussian random-walk proposal with adaptive covariance to update $\bm{X}$. The 5-dimensional vector of proposed values for $(\bm{X},(\sigma_t^j)^2)$ is then accepted or rejected according to the standard Metropolis-Hastings criterion \citep{metropolis1953equation, hastings1970monte}. The proposal for $\bm{X}$  follows the adaptive Metropolis (AM) approach of \cite{haario2001adaptive}, adjusting  the proposal covariance matrix based on all previous samples of $\bm{X}$:
\begin{equation}
C_{\ell}^*= s_d \, \text{Cov} (\bm{X}_0,\ldots, \bm{X}_{\ell}) + s_d \epsilon_0 I_d \, .
\label{eq:AMcov}
\end{equation}
Here $C_{\ell}^*$ is the proposal covariance matrix at step $\ell$, $I_d$ is the $d$-dimensional identity matrix, $\epsilon_0>0$ is a small constant to make $C_{\ell}^*$ positive definite, $d=4$ is the dimension of $\bm{X}$, and $s_{d} = 2.4^2/d$. The value of the scaling parameter $s_{d}$ is a standard choice to optimize the mixing properties of the Metropolis search \citep{gelman1996efficient}. This value might affect the efficiency of MCMC, but not the posterior distribution itself.

\section{Results and Discussion}
We apply the Bayesian method to all 16 sensors \added[id=CG]{at (1) cp = 10 MPa, ds = 6 MPa; (2) cp = 20 MPa, ds = 10 MPa; and (3) cp =30 MPa, ds = 10 MPa.} (sensor $13$ did not work during the experiment). For each sensor, we first calculate the parameter $\alpha^j$ using \eqref{eq:alphaj}. The values of $\alpha^j$ for the 16 sensors are shown in Table \ref{tab:para_post_1} \added[id=CG]{ -- \ref{tab:para_post_3}}. The sensors at the top half of the cylinder sample (sensors $16$, $4$, $12$, $8$, $15$, $3$, $11$, $7$), which are closer to the ball bouncing source, generally have lower $\alpha^j$ than sensors at the bottom half of the cylinder sample, e.g., sensors $6$, $14$, $2$, $10$, $5$, $13$, $1$, $9$. This is because the sensors close to the source have better signal-to-noise ratio than sensors away from the source; in other words, $\alpha^j$ is an indicator of signal quality.

\begin{figure}
\begin{center}
\includegraphics[scale=0.4]{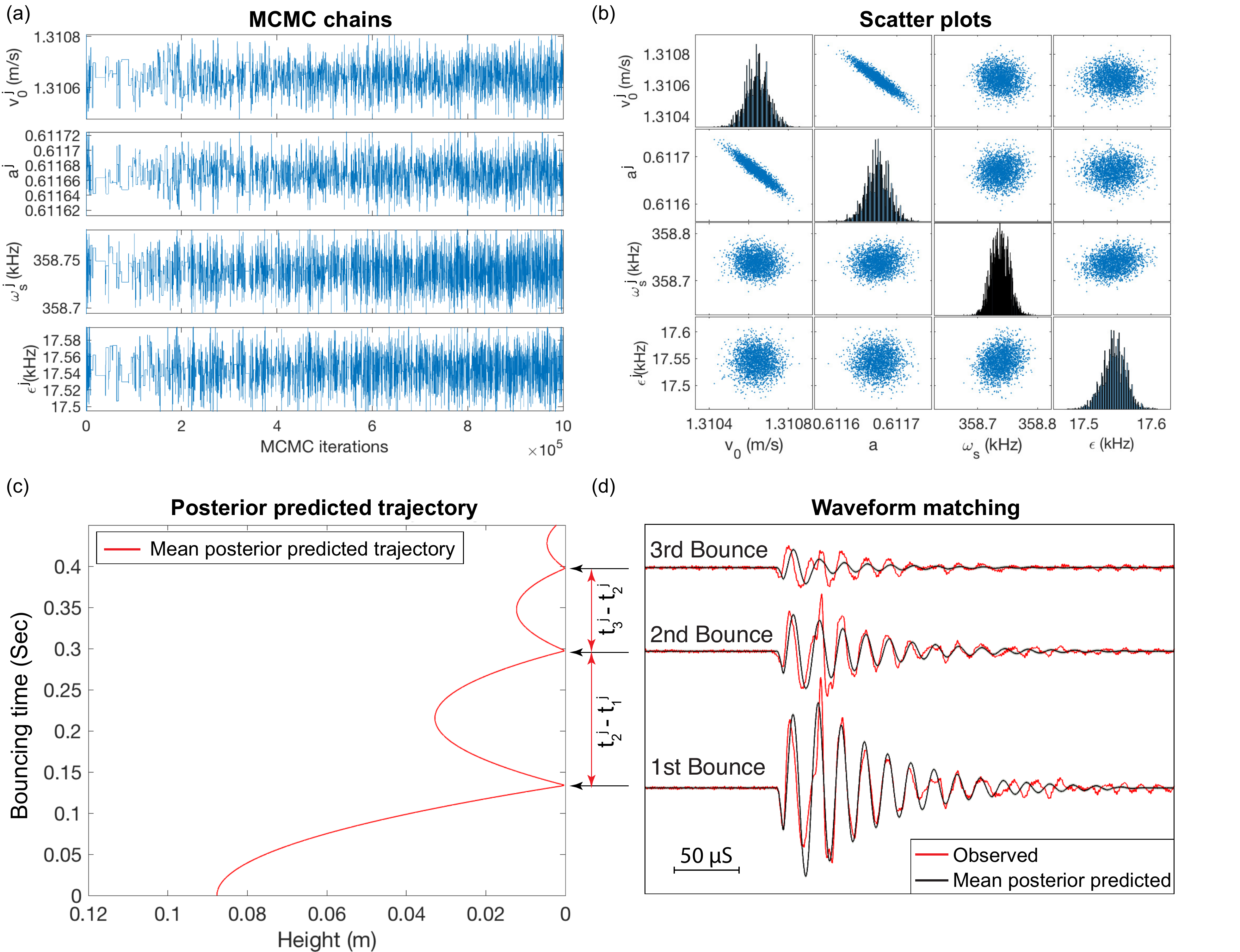}
\centering
\caption{MCMC chains and posterior distribution of four parameters for sensor $16$. The first 
$6\times10^5$ iterations of MCMC chains are discarded as burn-in. (a) MCMC chains for initial velocity $v_0^j$, rebound coefficient $a^j$, resonance frequency $\omega_s^j$, and damping coefficient $\varepsilon^j$. (b) Scatter plots of four parameters corresponding to MCMC chains. (c) Mean posterior predicted trajectory of ball bouncing. (d) Waveform comparison between observed (red) and mean posterior predicted waveforms (black).}
\label{fig:post_sensor_16}
\end{center}
\end{figure}

We perform $10^6$ MCMC iterations to explore the posterior \eqref{eqn:simplifiedposterior} for each sensor at varied pressures. The first $6\times10^5$ iterations of each MCMC chain are discarded as burn-in. We show MCMC chains and posterior distributions of $v_0^j$, $a^j$, $\omega_s^j$, and $\varepsilon^j$ for sensor $16$ \added[id=CG]{at cp = 30 MPa, ds = 10 MPa} in Figures~\ref{fig:post_sensor_16}(a) and (b). Figure~\ref{fig:post_sensor_16}(c) shows the mean posterior predicted trajectory of ball bouncing events. The comparison between the observed AE data and mean posterior predicted waveforms is shown in Figure~\ref{fig:post_sensor_16}(d). 

The marginal posterior distributions  of $(\sigma_t^j)^2$, $v_0^j$, $a^j$, $\omega_s^j$, and $\varepsilon^j$ \added[id=CG]{at (1) cp = 10 MPa, ds = 6 MPa; (2) cp = 20 MPa, ds = 10 MPa; and (3) cp =30 MPa, ds = 10 MPa} are summarized (via their means and standard deviations) in Table \ref{tab:para_post_1} \added[id=CG]{ - \ref{tab:para_post_3}}. The parameters $\omega_s^j$ and $\varepsilon^j$ for sensors closer to the bouncing source have higher posterior standard deviations than for sensors farther away from the bouncing source. 

The posterior distributions of $(\sigma_t^j)^2$, $v_0^j$, and $a^j$ are relatively similar across the sensors; this is expected, as the bouncing time data sets are the same for all sensors. Note that the posterior variance of $(\sigma_t^j)^2$ is roughly half of the prior variance, and that the mean of  $(\sigma_t^j)^2$ shifts slightly from its prior value.

\begin{figure}
\begin{center}
\includegraphics[scale=0.6]{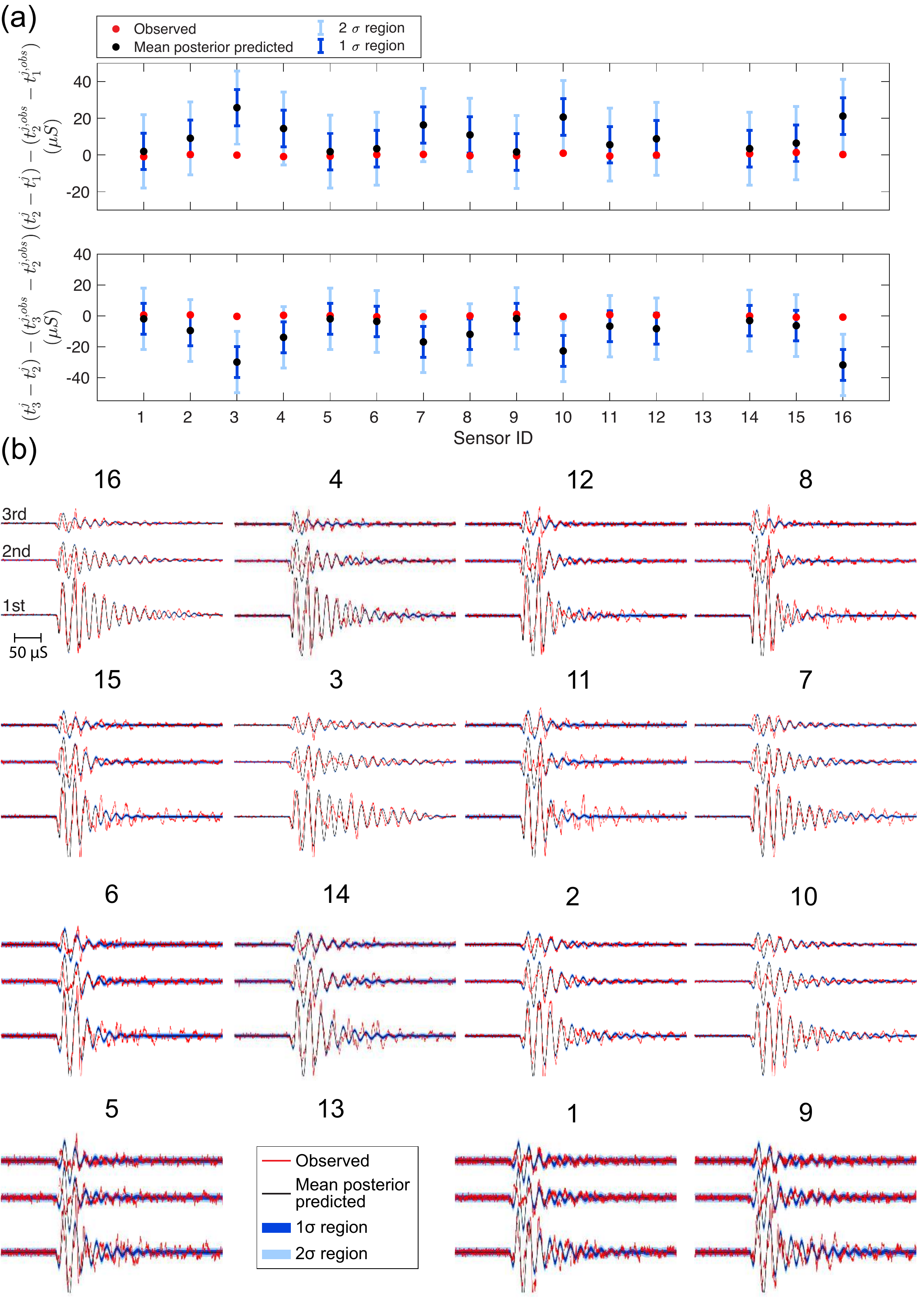}
\centering
\caption{Results at cp = 10 MPa, ds = 6 MPa: (a) Comparison between observed (red) and mean posterior predicted (blue) bouncing time intervals. The error bars indicate the 1$\sigma$ and 2$\sigma$ regions. (b) Waveform comparison between observed (red) and mean posterior predicted (black) waveforms of three bouncing events for 16 PZT sensors. Blue and light blue shading areas show the 1$\sigma$ and 2$\sigma$ regions of posterior predicted waveforms after the burn-in. The title of each subplot denotes sensor ID. Subplots are arranged in the order of sensor locations shown in Figure \ref{fig:sensor_distr}. Sensor $13$ did not work, so we put the legend in the position of sensor $13$.}
\label{fig:dt_wfm_16ch_cp10_ds6}
\end{center}
\end{figure}

\begin{figure}
\begin{center}
\includegraphics[scale=0.6]{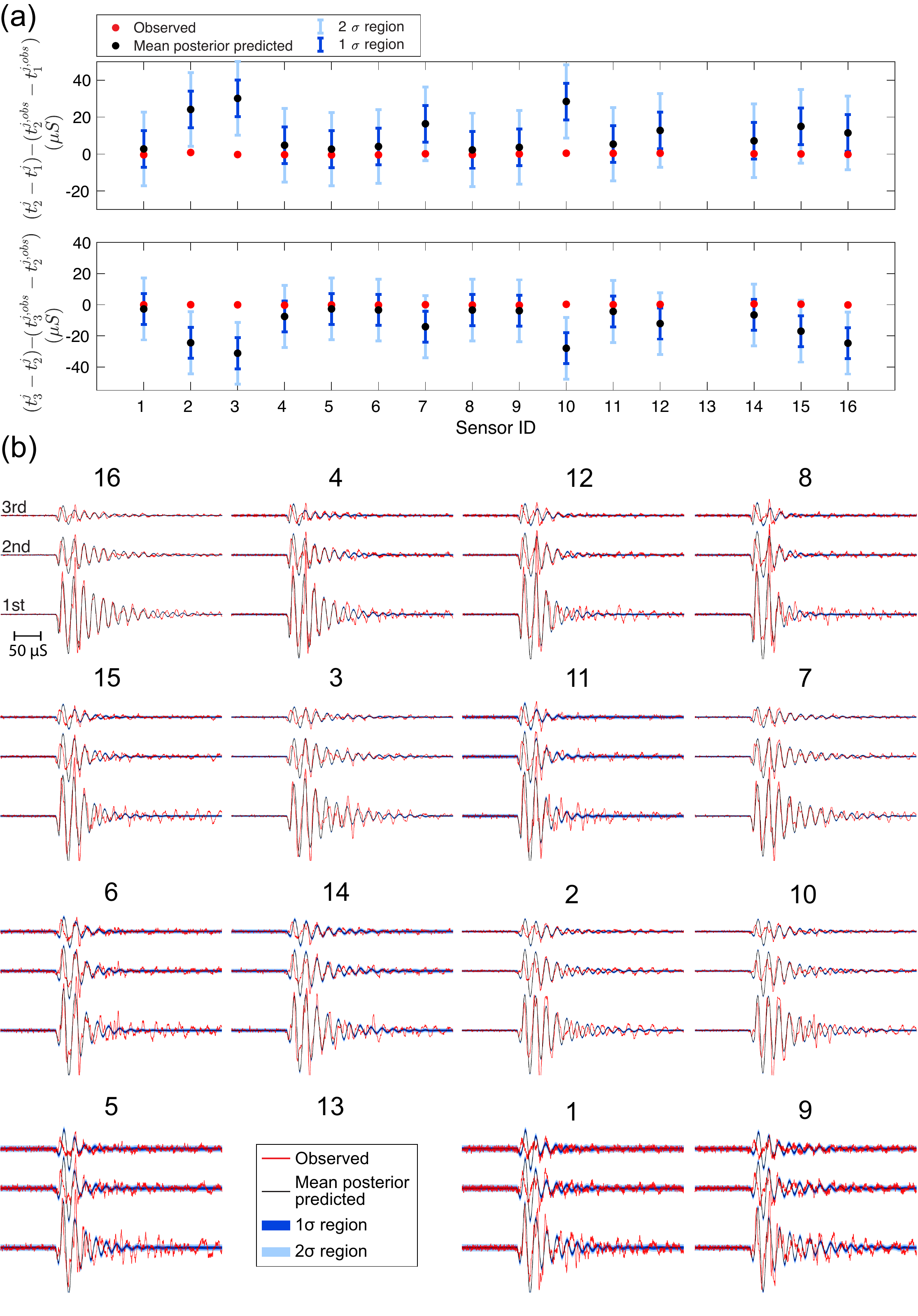}
\centering
\caption{Results at cp = 20 MPa, ds = 10 MPa: (a) Comparison between observed (red) and mean posterior predicted (blue) bouncing time intervals. The error bars indicate the 1$\sigma$ and 2$\sigma$ regions. (b) Waveform comparison between observed (red) and mean posterior predicted (black) waveforms of three bouncing events for 16 PZT sensors. Blue and light blue shading areas show the 1$\sigma$ and 2$\sigma$ regions of posterior predicted waveforms after the burn-in. The title of each subplot denotes sensor ID. Subplots are arranged in the order of sensor locations shown in Figure \ref{fig:sensor_distr}. Sensor $13$ did not work, so we put the legend in the position of sensor $13$.}
\label{fig:dt_wfm_16ch_cp20_ds10}
\end{center}
\end{figure}

\begin{figure}
\begin{center}
\includegraphics[scale=0.6]{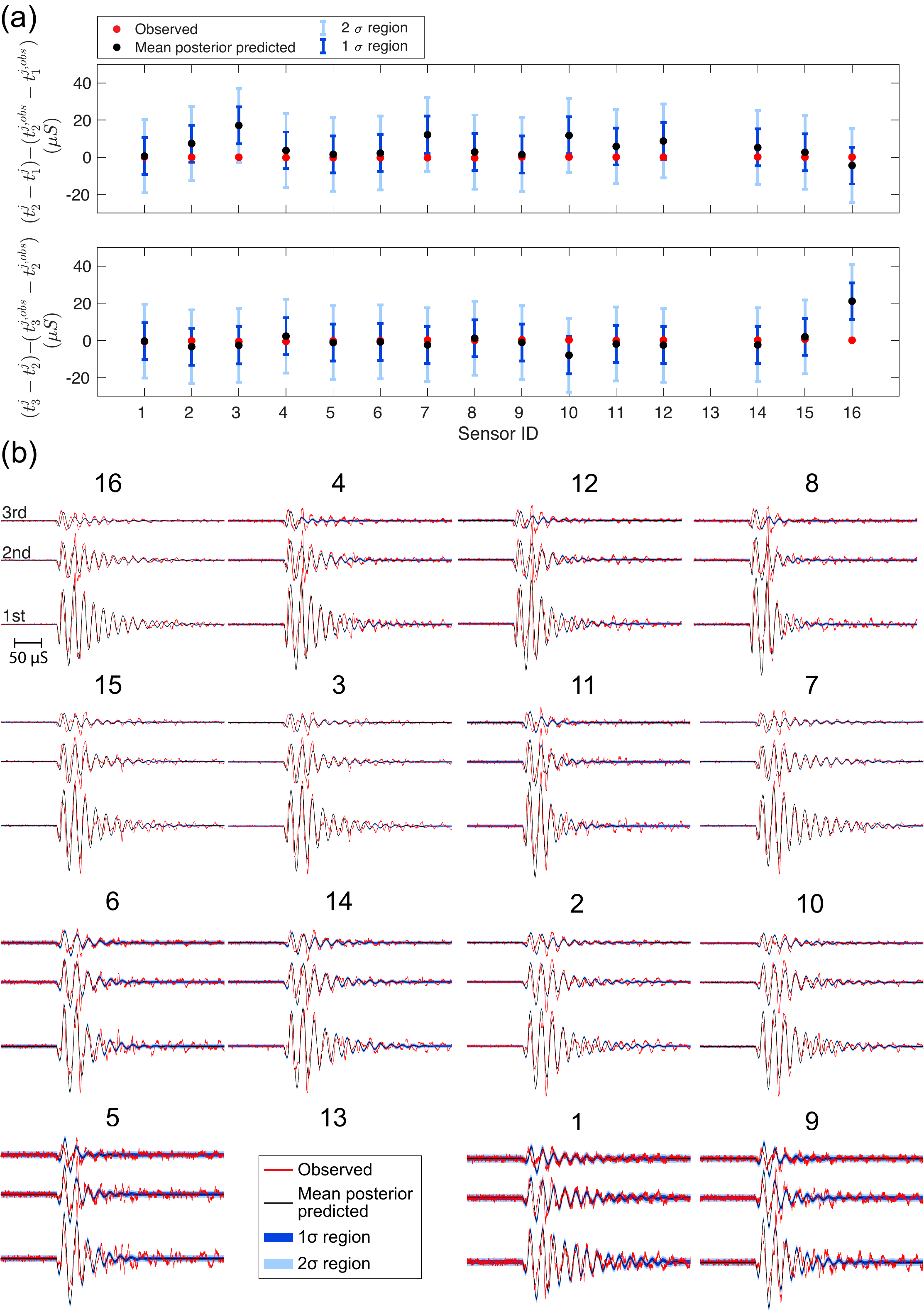}
\centering
\caption{Results at cp = 30 MPa, ds = 10 MPa: (a) Comparison between observed (red) and mean posterior predicted (blue) bouncing time intervals. The error bars indicate the 1$\sigma$ and 2$\sigma$ regions. (b) Waveform comparison between observed (red) and mean posterior predicted (black) waveforms of three bouncing events for 16 PZT sensors. Blue and light blue shading areas show the 1$\sigma$ and 2$\sigma$ regions of posterior predicted waveforms after the burn-in. The title of each subplot denotes sensor ID. Subplots are arranged in the order of sensor locations shown in Figure \ref{fig:sensor_distr}. Sensor $13$ did not work, so we put the legend in the position of sensor $13$.}
\label{fig:dt_wfm_16ch_cp30_ds10}
\end{center}
\end{figure}

\replaced[id=CG]{Figure \ref{fig:dt_wfm_16ch_cp10_ds6} (a) , Figure \ref{fig:dt_wfm_16ch_cp20_ds10} (a) and , Figure \ref{fig:dt_wfm_16ch_cp30_ds10} (a) show}{Figure 6 (a) shows} the comparison between observed and mean posterior predicted bouncing time intervals, $t_2^j - t_1^j$ and $t_3^j - t_2^j$, for all the sensors \added[id=CG]{at (1) cp = 10 MPa, ds = 6 MPa; (2) cp = 20 MPa, ds = 10 MPa; and (3) cp =30 MPa, ds = 10 MPa}. The bias is smaller than 20 $\mu$s. \replaced[id=CG]{Figure \ref{fig:dt_wfm_16ch_cp10_ds6} (b) , Figure \ref{fig:dt_wfm_16ch_cp20_ds10} (b) and , Figure \ref{fig:dt_wfm_16ch_cp30_ds10} (b) show}{Figure 6 (b) shows} the comparison between observed and mean posterior predicted waveforms. The observed waveforms are all well predicted. Blue and light blue shaded areas show the 1-$\sigma$ and 2-$\sigma$ regions of the posterior predictive waveforms (marginal intervals at each timestep). 

The 2-$\sigma$ region of the posterior predictive waveforms (light blue shadow areas) almost covers the observed waveforms. The higher the noise levels of the observations, the larger the light blue shadow areas. In contrast, the sensors with high signal quality generally show larger bias in bouncing time intervals. This is probably because of the tradeoff between the likelihood functions $P(\bm{\delta t}|v_0^j, a^j)$ and $P(\bm{d}^j|v_0^j, a^j, \omega_s^j, \varepsilon^j)$.

\begin{figure}
\begin{center}
\includegraphics[scale=0.45]{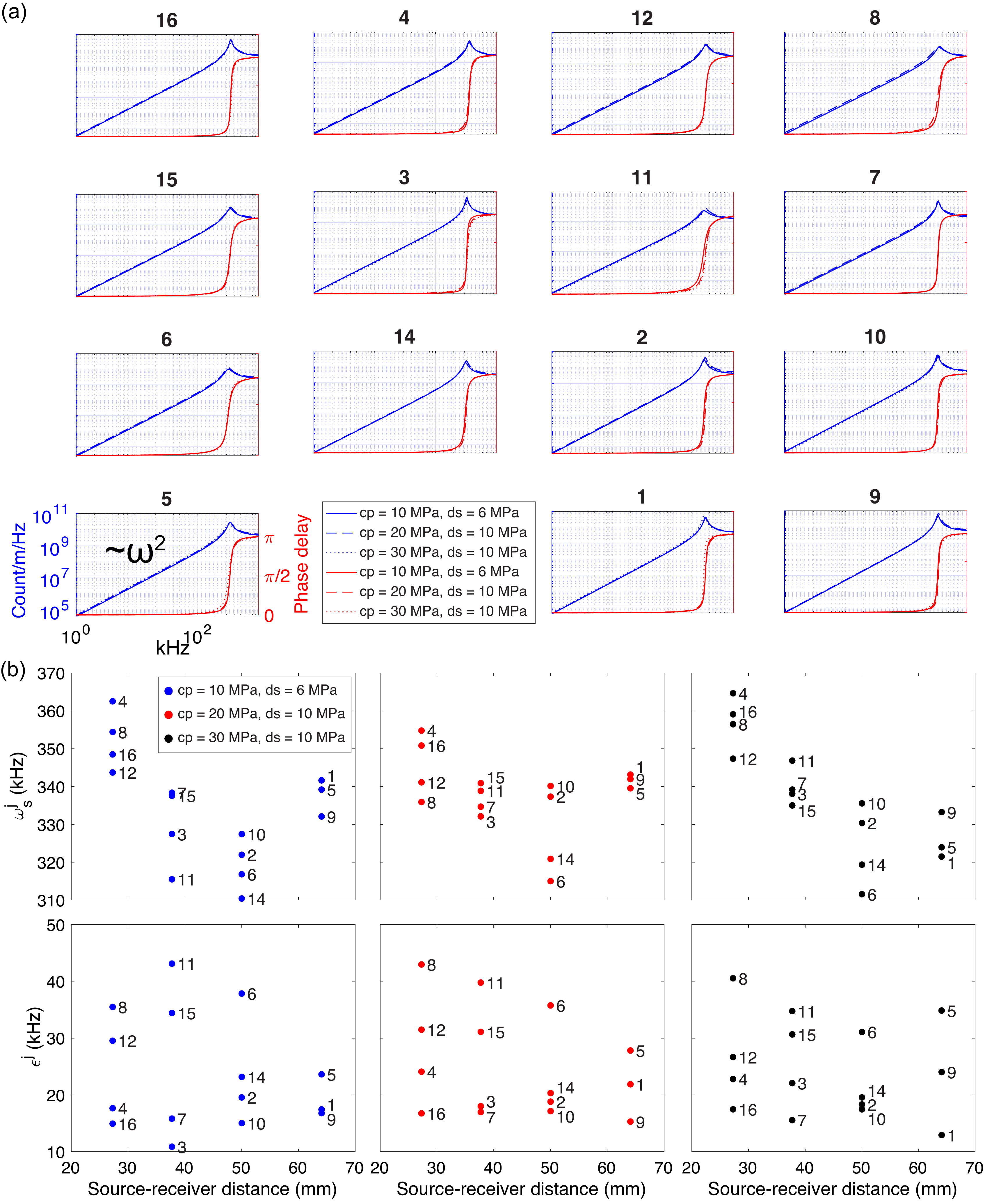}
\centering
\caption{\added[id=CG]{Transfer functions at different cp and ds.} (a) Mean posterior predicted amplitude response (blue) and phase delay (red). The amplitude response tends to a constant at high frequencies, and is proportional to $\omega^2$ at low frequencies. The title of each subplot denotes sensor ID. Subplots are arranged in the order of sensor locations shown in Figure \ref{fig:sensor_distr}. Sensor $13$ did not work, so we put the legend in the position of sensor $13$. (b) $\omega_s^j$ and $\varepsilon^j$ as a function of source-receiver distance.}
\label{fig:rsps}
\end{center}
\end{figure}

The mean posterior resonance frequency $\omega_s^j$ for all the sensors varies from \replaced[id=CG]{310}{311} to  \replaced[id=CG]{365}{364} kHz, and the damping coefficient $\varepsilon^j$ varies from \replaced[id=CG]{11}{13} to \replaced[id=CG]{43}{40} kHz. The posterior standard deviations for $\omega_s^j$ and $\varepsilon^j$ are all within 1 kHz. With the posterior distributions of $\omega_s^j$ and $\varepsilon^j$, we can obtain frequency-response functions of all sensors using \eqref{eq:resp}(a). The standard deviation indicates how reliable the response function of each sensor is. We show the mean posterior amplitude response and phase delay of all response functions \added[id=CG]{at varied cp and ds} in Figure \ref{fig:rsps}(a). The amplitude response tends to a constant at high frequencies, and is proportional to $\omega^2$ at low frequencies. \added[id=CG]{The amplitude response at higher cp is larger than that at lower cp, but the difference is not obvious.} The phase delay is close to zero at low frequencies and tends to $\pi$ at high frequencies. The \textit{in situ} response functions can be used to calibrate real AE data, i.e., convert digital AE data into time series with physical units, for fracturing experiments in rocks under high pressure conditions.

We plot $\omega_s^j$ and $\varepsilon^j$ as a function of source-receiver distance in Figure \ref{fig:rsps}(b). $\omega_s^j$ shows a clear trend of decay with the increasing source-receiver distance, indicating that attenuation effects, which are not included in our model, should be taken into account in \eqref{eq:3C_disp} to avoid mapping sample $Q$ into instrument response functions. $\varepsilon^j$ does not show any distance-dependent properties. 

\begin{table}
    \centering
    \caption{Posterior mean and standard deviation (std dev) of $(\sigma_t^j)^2$, $v_0^j$, $a^j$, $\omega_s^j$, and $\varepsilon^j$ for 16 sensors \added[id=CG]{at cp = 10 MPa and ds = 6 MPa}.}
    \begin{tabular}{l|c|c|c|c|c|c|c|c|c|c|c}
    \hline
         \multirow{2}{*}{ID} & \multirow{2}{*}{$\alpha^j$} & \multicolumn{2}{c|}{$(\sigma_t^j)^2$ ($10^{-10} s^2$)} & \multicolumn{2}{c|}{$v_0^j$ (m/s)} & \multicolumn{2}{c|}{$a^j$} & \multicolumn{2}{c|}{$\omega_s^j$ (kHz)} &\multicolumn{2}{c|}{$\varepsilon^j$ (kHz)}\\
                                    & & mean & std dev & mean & std dev & mean & std dev & mean & std dev & mean & std dev \\
         \hline
1 & 0.144 & 0.918 & 0.074 & 1.02E+00 & 4.82E-05 & 6.74E-01 & 2.29E-05 &  341.63 & 0.19 & 17.44 & 0.12\\ 
2 & 0.027 & 0.918 & 0.074 & 1.02E+00 & 4.79E-05 & 6.74E-01 & 2.28E-05 &  321.97 & 0.05 & 19.55 & 0.05\\ 
3 & 0.007 & 0.919 & 0.075 & 1.02E+00 & 4.80E-05 & 6.74E-01 & 2.28E-05 &  327.47 & 0.01 & 10.88 & 0.02\\ 
4 & 0.019 & 0.915 & 0.073 & 1.02E+00 & 4.81E-05 & 6.74E-01 & 2.28E-05 &  362.52 & 0.05 & 17.66 & 0.04\\ 
5 & 0.128 & 0.918 & 0.074 & 1.02E+00 & 4.79E-05 & 6.74E-01 & 2.28E-05 &  339.19 & 0.27 & 23.63 & 0.18\\ 
6 & 0.059 & 0.918 & 0.074 & 1.02E+00 & 4.80E-05 & 6.74E-01 & 2.29E-05 &  316.82 & 0.15 & 37.85 & 0.17\\ 
7 & 0.012 & 0.916 & 0.074 & 1.02E+00 & 4.78E-05 & 6.74E-01 & 2.27E-05 &  338.36 & 0.03 & 15.83 & 0.03\\ 
8 & 0.019 & 0.917 & 0.074 & 1.02E+00 & 4.77E-05 & 6.74E-01 & 2.27E-05 &  354.41 & 0.08 & 35.49 & 0.10\\ 
9 & 0.137 & 0.918 & 0.074 & 1.02E+00 & 4.79E-05 & 6.74E-01 & 2.28E-05 &  332.08 & 0.18 & 16.80 & 0.16\\ 
10 & 0.015 & 0.918 & 0.074 & 1.02E+00 & 4.80E-05 & 6.74E-01 & 2.28E-05 &  327.44 & 0.03 & 15.03 & 0.03\\ 
11 & 0.028 & 0.918 & 0.074 & 1.02E+00 & 4.82E-05 & 6.74E-01 & 2.30E-05 &  315.49 & 0.10 & 43.12 & 0.14\\ 
12 & 0.025 & 0.917 & 0.074 & 1.02E+00 & 4.78E-05 & 6.74E-01 & 2.28E-05 &  343.71 & 0.09 & 29.54 & 0.09\\ 
13 & - & - & - & - & - & - & - & - & - & - & -\\ 
14 & 0.044 & 0.918 & 0.074 & 1.02E+00 & 4.81E-05 & 6.74E-01 & 2.29E-05 &  310.41 & 0.07 & 23.17 & 0.08\\ 
15 & 0.023 & 0.918 & 0.075 & 1.02E+00 & 4.81E-05 & 6.74E-01 & 2.30E-05 &  337.57 & 0.08 & 34.43 & 0.10\\ 
16 & 0.006 & 0.922 & 0.073 & 1.02E+00 & 4.87E-05 & 6.74E-01 & 2.35E-05 &  348.49 & 0.02 & 14.93 & 0.02\\ 
    \hline
    \end{tabular}
    \label{tab:para_post_1}
\end{table}

\begin{table}
    \centering
    \caption{Posterior mean and standard deviation (std dev) of $(\sigma_t^j)^2$, $v_0^j$, $a^j$, $\omega_s^j$, and $\varepsilon^j$ for 16 sensors \added[id=CG]{at cp = 20 MPa and ds = 10 MPa}.}
    \begin{tabular}{l|c|c|c|c|c|c|c|c|c|c|c}
    \hline
         \multirow{2}{*}{ID} & \multirow{2}{*}{$\alpha^j$} & \multicolumn{2}{c|}{$(\sigma_t^j)^2$ ($10^{-10} s^2$)} & \multicolumn{2}{c|}{$v_0^j$ (m/s)} & \multicolumn{2}{c|}{$a^j$} & \multicolumn{2}{c|}{$\omega_s^j$ (kHz)} &\multicolumn{2}{c|}{$\varepsilon^j$ (kHz)}\\
                                    & & mean & std dev & mean & std dev & mean & std dev & mean & std dev & mean & std dev \\
         \hline
1 & 0.082 & 0.918 & 0.074 & 1.13E+00 & 4.89E-05 & 6.64E-01 & 2.08E-05 &  343.14 & 0.29 & 21.87 & 0.12\\ 
2 & 0.011 & 0.917 & 0.074 & 1.13E+00 & 4.90E-05 & 6.64E-01 & 2.09E-05 &  337.33 & 0.04 & 18.80 & 0.03\\ 
3 & 0.007 & 0.918 & 0.074 & 1.13E+00 & 4.88E-05 & 6.63E-01 & 2.08E-05 &  332.10 & 0.02 & 18.01 & 0.03\\ 
4 & 0.010 & 0.917 & 0.074 & 1.13E+00 & 4.92E-05 & 6.64E-01 & 2.08E-05 &  354.77 & 0.04 & 24.08 & 0.04\\ 
5 & 0.070 & 0.918 & 0.074 & 1.13E+00 & 4.91E-05 & 6.64E-01 & 2.08E-05 &  339.54 & 0.18 & 27.82 & 0.16\\ 
6 & 0.026 & 0.918 & 0.075 & 1.13E+00 & 4.89E-05 & 6.64E-01 & 2.08E-05 &  315.00 & 0.09 & 35.76 & 0.11\\ 
7 & 0.006 & 0.918 & 0.074 & 1.13E+00 & 4.92E-05 & 6.64E-01 & 2.09E-05 &  334.67 & 0.02 & 16.98 & 0.02\\ 
8 & 0.012 & 0.918 & 0.074 & 1.13E+00 & 4.89E-05 & 6.64E-01 & 2.08E-05 &  335.91 & 0.07 & 42.97 & 0.08\\ 
9 & 0.061 & 0.918 & 0.074 & 1.13E+00 & 4.88E-05 & 6.64E-01 & 2.08E-05 &  341.97 & 0.19 & 15.30 & 0.13\\ 
10 & 0.010 & 0.917 & 0.074 & 1.13E+00 & 4.93E-05 & 6.63E-01 & 2.09E-05 &  340.14 & 0.04 & 17.14 & 0.03\\ 
11 & 0.026 & 0.918 & 0.075 & 1.13E+00 & 4.92E-05 & 6.64E-01 & 2.09E-05 &  338.87 & 0.11 & 39.78 & 0.13\\ 
12 & 0.011 & 0.918 & 0.074 & 1.13E+00 & 4.91E-05 & 6.64E-01 & 2.08E-05 &  341.11 & 0.06 & 31.49 & 0.07\\ 
13 & - & - & - & - & - & - & - & - & - & - & -\\ 
14 & 0.029 & 0.917 & 0.074 & 1.13E+00 & 4.94E-05 & 6.64E-01 & 2.09E-05 &  320.88 & 0.06 & 20.32 & 0.06\\ 
15 & 0.011 & 0.918 & 0.074 & 1.13E+00 & 4.89E-05 & 6.64E-01 & 2.07E-05 &  340.89 & 0.05 & 31.11 & 0.06\\ 
16 & 0.003 & 0.917 & 0.073 & 1.13E+00 & 4.93E-05 & 6.64E-01 & 2.08E-05 &  350.82 & 0.01 & 16.76 & 0.01\\ 
    \hline
    \end{tabular}
    \label{tab:para_post_2}
\end{table}

\begin{table}
    \centering
    \caption{Posterior mean and standard deviation (std dev) of $(\sigma_t^j)^2$, $v_0^j$, $a^j$, $\omega_s^j$, and $\varepsilon^j$ for 16 sensors \added[id=CG]{at cp = 30 MPa and ds = 10 MPa}.}
    \begin{tabular}{l|c|c|c|c|c|c|c|c|c|c|c}
    \hline
         \multirow{2}{*}{ID} & \multirow{2}{*}{$\alpha^j$} & \multicolumn{2}{c|}{$(\sigma_t^j)^2$ ($10^{-10} s^2$)} & \multicolumn{2}{c|}{$v_0^j$ (m/s)} & \multicolumn{2}{c|}{$a^j$} & \multicolumn{2}{c|}{$\omega_s^j$ (kHz)} &\multicolumn{2}{c|}{$\varepsilon^j$ (kHz)}\\
                                    & & mean & std dev & mean & std dev & mean & std dev & mean & std dev & mean & std dev \\
         \hline
        1 & 0.103 & 0.917 & 0.074 & 1.31100 & 5.42E-05 & 0.61152 & 1.88E-05 &  321.48 & 0.15 & 12.93 & 0.11\\ 
2 & 0.013 & 0.918 & 0.074 & 1.31115 & 5.42E-05 & 0.61147 & 1.87E-05 &  330.47 & 0.03 & 18.26 & 0.03\\ 
3 & 0.005 & 0.917 & 0.074 & 1.31130 & 5.47E-05 & 0.61144 & 1.89E-05 &  338.42 & 0.02 & 22.00 & 0.03\\ 
4 & 0.012 & 0.916 & 0.074 & 1.31102 & 5.48E-05 & 0.61152 & 1.90E-05 &  364.67 & 0.04 & 22.77 & 0.04\\ 
5 & 0.074 & 0.917 & 0.074 & 1.31103 & 5.44E-05 & 0.61151 & 1.89E-05 &  323.93 & 0.17 & 34.76 & 0.17\\ 
6 & 0.030 & 0.918 & 0.074 & 1.31104 & 5.46E-05 & 0.61151 & 1.90E-05 &  311.61 & 0.08 & 31.06 & 0.10\\ 
7 & 0.006 & 0.918 & 0.075 & 1.31122 & 5.53E-05 & 0.61146 & 1.91E-05 &  339.36 & 0.02 & 15.50 & 0.02\\ 
8 & 0.014 & 0.918 & 0.074 & 1.31102 & 5.45E-05 & 0.61152 & 1.89E-05 &  356.53 & 0.08 & 40.51 & 0.09\\ 
9 & 0.085 & 0.918 & 0.074 & 1.31103 & 5.48E-05 & 0.61151 & 1.90E-05 &  333.22 & 0.11 & 23.96 & 0.14\\ 
10 & 0.012 & 0.918 & 0.074 & 1.31128 & 5.44E-05 & 0.61143 & 1.88E-05 &  335.74 & 0.03 & 17.36 & 0.03\\ 
11 & 0.016 & 0.918 & 0.074 & 1.31111 & 5.45E-05 & 0.61149 & 1.89E-05 &  347.06 & 0.07 & 34.66 & 0.08\\ 
12 & 0.013 & 0.917 & 0.074 & 1.31116 & 5.47E-05 & 0.61147 & 1.90E-05 &  347.68 & 0.06 & 26.62 & 0.06\\ 
13 & - & - & - & - & - & - & - & - & - & - & -\\ 
14 & 0.021 & 0.919 & 0.074 & 1.31110 & 5.49E-05 & 0.61149 & 1.91E-05 &  319.47 & 0.05 & 19.52 & 0.05\\ 
15 & 0.012 & 0.918 & 0.074 & 1.31101 & 5.43E-05 & 0.61152 & 1.88E-05 &  335.04 & 0.05 & 30.65 & 0.05\\ 
16 & 0.004 & 0.917 & 0.073 & 1.31064 & 5.46E-05 & 0.61167 & 1.87E-05 &  358.74 & 0.01 & 17.55 & 0.02\\ 
    \hline
    \end{tabular}
    \label{tab:para_post_3}
\end{table}

\section{Conclusion}
We develop a Bayesian waveform-based method to calibrate PZT sensors of a newly designed \textit{in situ} ball drop system in a sealed pressure vessel. Taking full waveforms due to ball bounces as input data, the Bayesian method successfully infers the model parameters $v_0^j$, $a^j$, $\omega_s^j$, and $\varepsilon^j$. Both the posterior distributions of \textit{in situ} response functions of PZT sensors and the trajectories of ball bounces are recovered by this method.

With the \textit{in situ} estimation of frequency-dependent sensor response functions, we are able to convert the AE waveforms' amplitude and phase to real physical parameters (e.g., displacements or accelerations) under high pressure conditions. The obtained uncertainties of response functions indicate the reliability of each sensor.

Our proposed method was tested on a titanium cylinder with a very homogeneous structure. For more complex (and realistic) cases, additional work needs to be performed. A good estimate of wave speeds is required, and, for example, attenuation in other rock types can be significant \citep{lockner1977changes,winkler1979friction}. As shown in Figure \ref{fig:rsps}(b), attenuation may be mapped into instrument response if not accounted for. We believe that using multiple bounces, as we have done here, will allow for a better constraint of the attenuation of the sample as well as for estimating wave speeds using relative arrival times and cross-correlation methods \citep{waldhauser2000double,zhang2003double,fuenzalida2013high,weemstra2013seismic}.

\begin{acknowledgments}
This research was supported by Total, Shell Global, and MIT Earth Resources Laboratory.
\end{acknowledgments}

\newpage
\bibliographystyle{gji}
\bibliography{Balldrop}

\appendix
\section{Waveforms after the 3rd bouncing event}
\label{sec:bc}
\begin{figure}
\begin{center}
\includegraphics[scale=0.45]{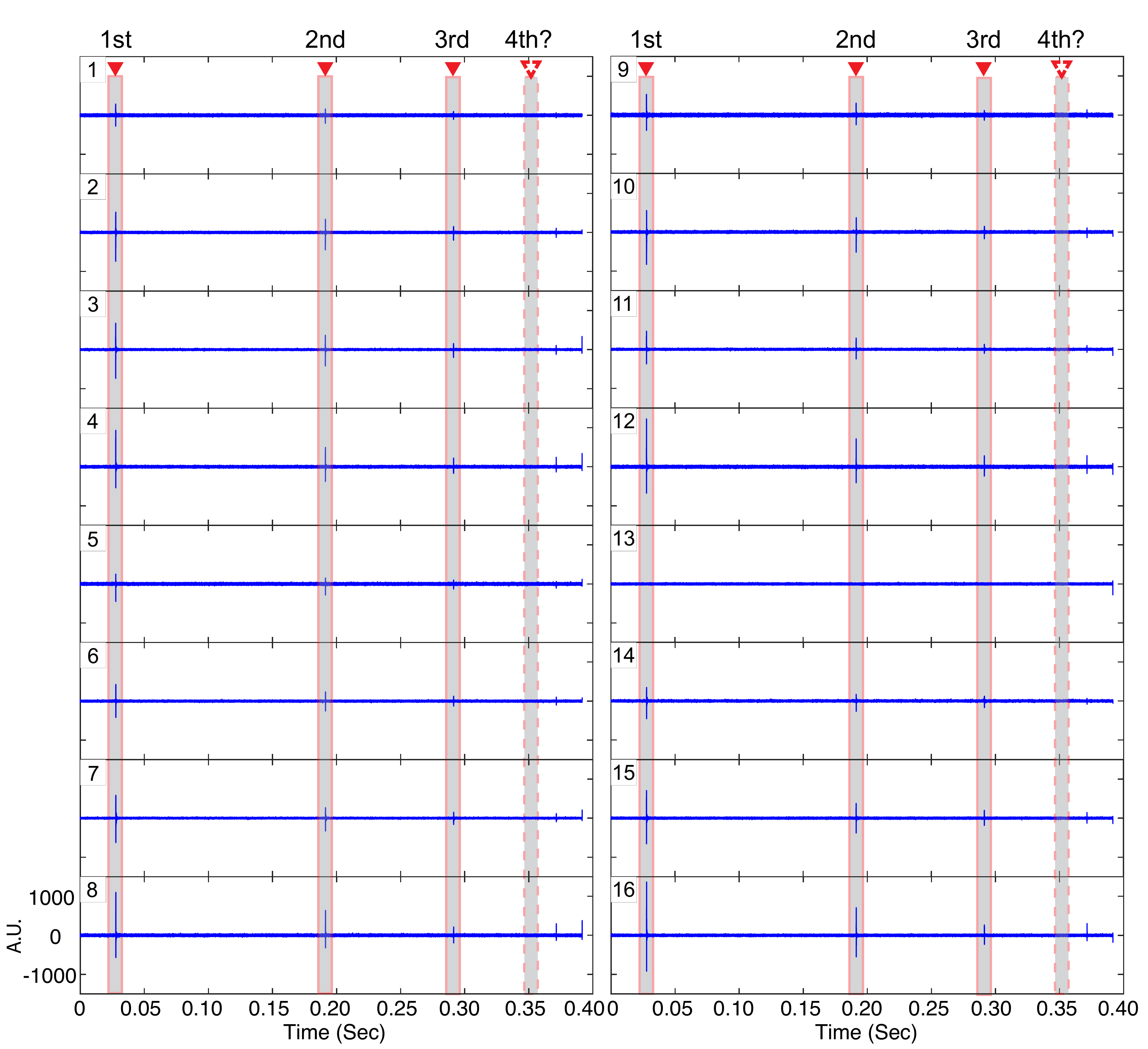}
\centering
\caption{Complete continuous waveforms containing waveforms after the 3rd bounce event for 16 sensors. The first three bouncing events are denoted as solid red triangles. The dashed triangle marks the theoretical arrival time of bounces based on Equation~\ref{eq:dt}.}
\label{fig:wfm_after_3rd}
\end{center}
\end{figure}

In the main text, we only use waveform data for the first three bounces. We show the complete continuous waveforms containing waveforms after the third bounce event for 16 sensors in Figure~\ref{fig:wfm_after_3rd}. The expected fourth bouncing event, marked as a dashed triangle, does not appear around the theoretical time, but around 0.2 sec later. The fourth bouncing event even presents higher amplitude than the third bouncing event at sensor 4, 8, 12 and 16. This indicates that after the third bounce, when the rebound vertical velocity becomes 21.6\% of the initial velocity $v_0$ and the maximum rebound height becomes 4.9 mm (comparable to the radius of the ball 3.18 mm), the simple rebound model cannot predict the ball's motion. The inclusion of other forces neglected in the main text, e.g., the drag force due to air resistance, and the Magnus force due to the ball's spin, and the buoyant force,  may help to improve the simple rebound model and predict the ball's trajectory after the third bounce; however, that is beyond the scope of this paper. 

\section{Directional response of a sensor}
\label{sec:amp_inc}
\added[id=CG]{We justify the cosine approximation of the directional response of a sensor in Equation~\ref{eqn:directional} in this appendix. \citet{tang1994radiation} derived the radiation patterns of an elastic wave field generated by circular plane compressional and shear transducers. The amplitudes of generated compressional waves due to a compressional PZT sensor at a distance of $R$ away from the sensor and an angle $\theta$ with the normal direction of the sensor surface is}

\begin{equation}
\begin{split}
    u_R =& \frac{R_s^2 \sigma_{zz}}{4\pi \mu} \frac{exp(-ik_\alpha R)}{R}\left[\frac{J_1(k_\alpha R_s\sin \theta)}{k_\alpha R_s\sin \theta}\right]\\
    &\frac{(V_S/V_P)^2\cos \theta[1 - 2 (V_S/V_P)^2 \sin^2 \theta]}{[1-2(V_S/V_P)^2\sin^2\theta]^2 + 4(V_S/V_P)^3\sin^2\theta \cos \theta [1-(V_S/V_P)^2 \sin^2 \theta]^{1/2}},
\end{split}
\end{equation}
\added[id=CG]{where $R_s = 5 \text{mm}$ is the radius of the PZT sensor, $\sigma_{zz}$ is stress uniformly distributed on the surface of the piston surface of the PZT sensor, $V_P=6011.6~\text{m}/\text{s}$ and $V_S=3093.0~\text{m}/\text{s}$ are the compressional wave velocity and shear wave velocity of the titanium sample of the compressional waves, $k_\alpha$ is the wave number of P wave.}

\begin{figure}
    \centering
    \includegraphics[scale=0.4]{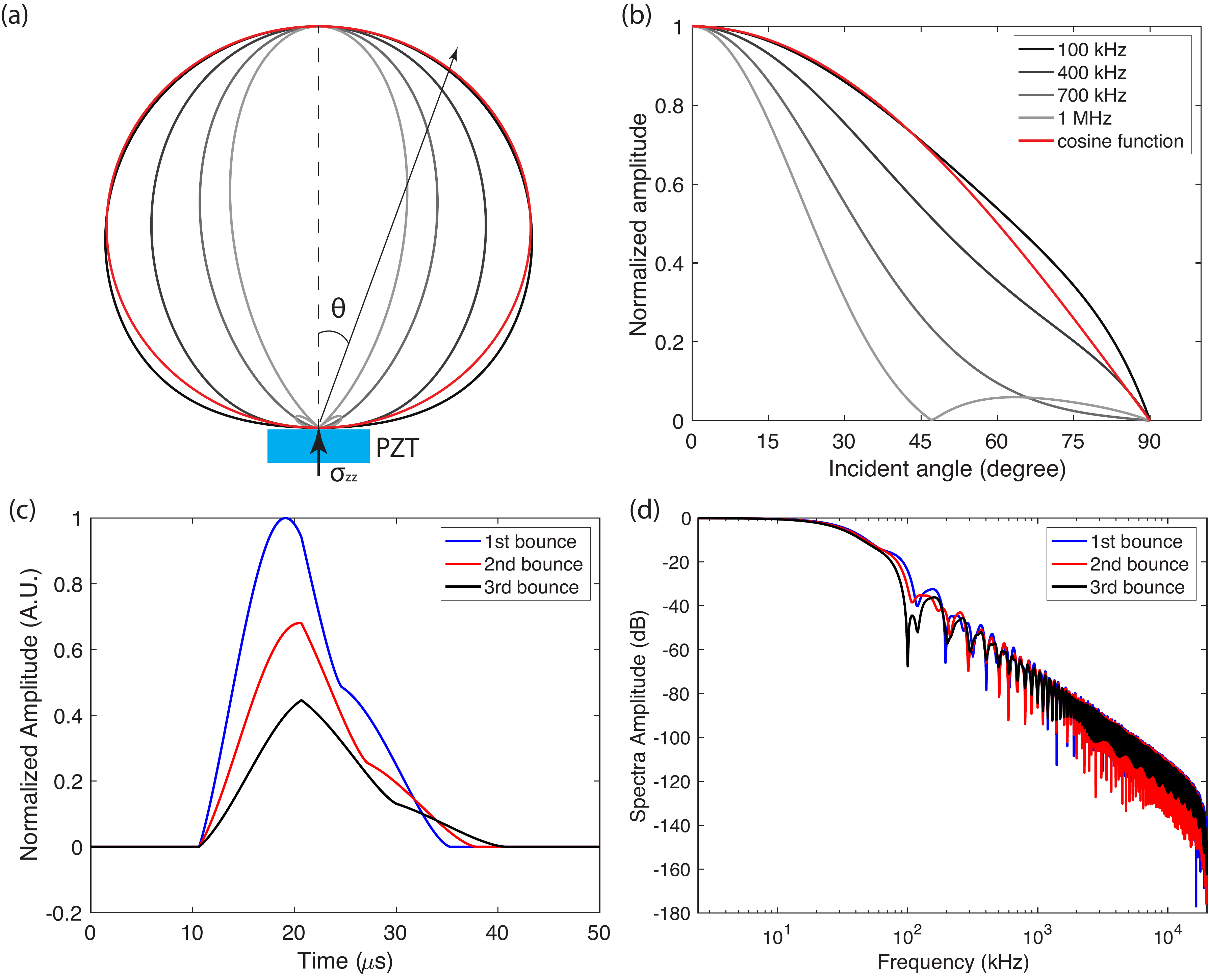}
    \caption{(a) The radiation pattern of compressional waves of a compressional PZT sensor at frequencies of 100 kHz, 400 kHz, 700 kHz, and 1 MHz. The red line denotes the radiation pattern of the cosine approximation. (b) The received amplitude as a function of incident angle. The red line denotes the radiation pattern of the cosine approximation. (c) The normalized compressional waveforms due to ball bounces at sensor 1. (d) The normalized spectra amplitude of compressional waveforms due to ball bounces at sensor 1.}
    \label{fig:directivity}
\end{figure}

\added[id=CG]{If we assume the directional response of a sensor as a transmitter and receiver are equivalent, the theoretical radiation pattern in \citet{tang1994radiation} can be used to model the received amplitude as a function of incident angle. We compared the theoretical radiation pattern of compressional waves and the received amplitude as a function incident angle $\theta$ at frequencies of 100 kHz, 400 kHz, 700 kHz, and 1 MHz in Figure \ref{fig:directivity} (a) and (b). The cosine dependence used in our paper is close to the theoretical solution at low frequency ($f = 100~\text{kHz}$). \citet{kwiatek2014improved} have used a bell-shaped function $h = exp(-a\alpha^b)$ to model the angular dependence, which is close to the theoretical pattern at medium frequency ($f = 700~\text{kHz}$). More complex multi-lobe patterns appear at high frequency ($f = 1~\text{MHz}$). Because the dominant frequency of waveforms due to ball bounces is less than 100 kHz (Figure \ref{fig:directivity}(c) and (d)), it is proper to use a cosine approximation.}

\section{Boundary effects on wave propagation modeling}
\label{sec:bd_wave}

\added[id=CG]{The Green's function in the main text is for an infinite unbounded homogeneous medium. To show that this approximation is valid for our experiment setup, we model the wave propagation in a more realistic environment using finite difference (FD) method with cylindrical coordinates. A cylindrical finite difference program is used to model the wave propagation in this ball drop system \citep{chen1998three}.}

\begin{figure}
    \centering
    \includegraphics[scale=0.25]{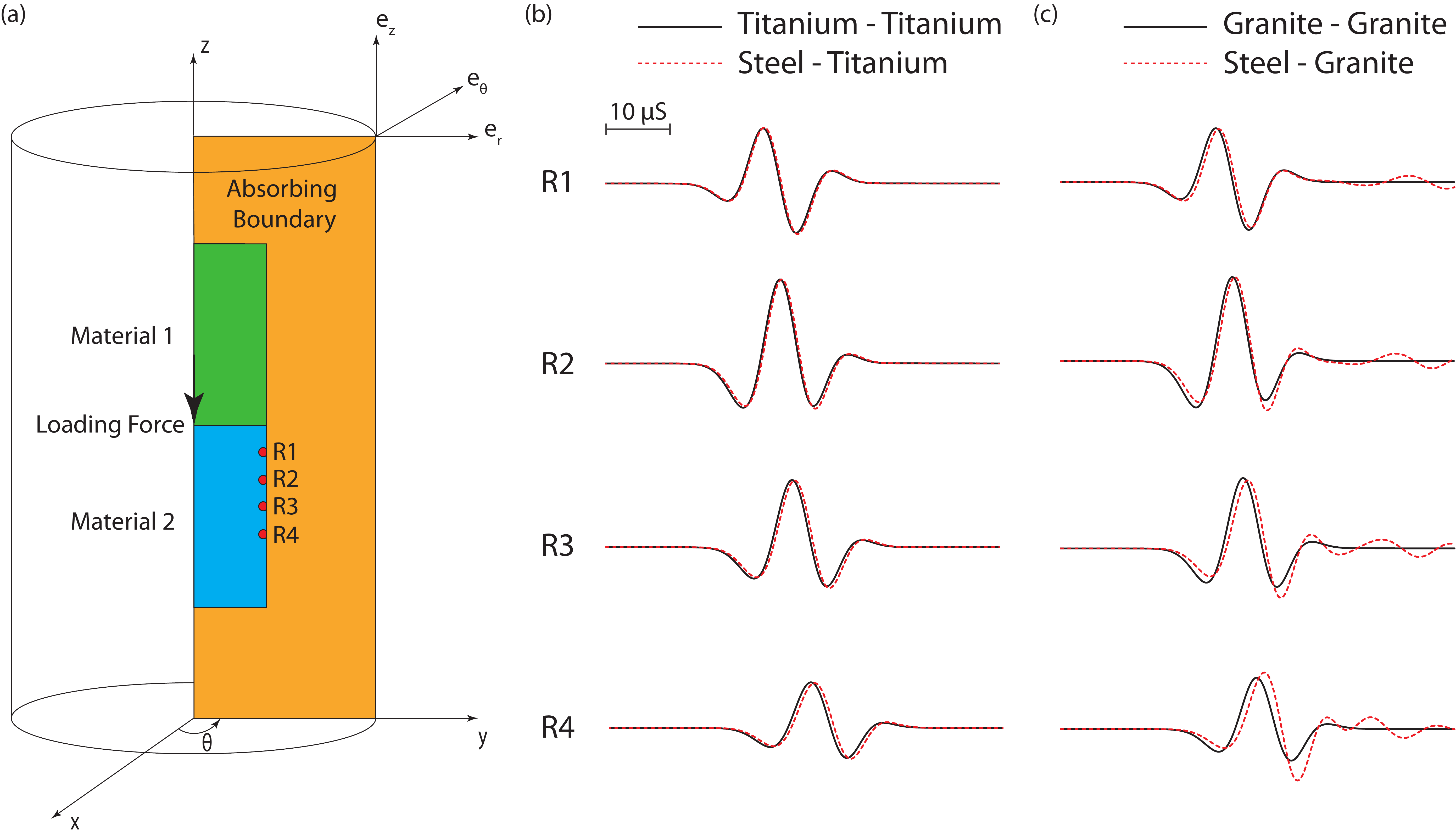}
    \caption{(a) A schematic of the finite difference model. The waveform comparison between the experimental setup with the ball drop apparatus made of Steel and the infinite unbounded homogeneous approximation at four receivers for (b) Titanium and (c) Granite samples.}
    \label{fig:fd}
\end{figure}

\added[id=CG]{Figure \ref{fig:fd}(a) shows the schematic of the FD model. A cross-section of the cylinder is plotted in color. The green region denotes material 1, which is the ball drop apparatus. The blue region denotes material 2, which is the sample. The four red circles show the locations of sensors, which are the same locations relative to the loading force as sensors in the experiment. The loading force function is set to be a Ricker wavelet with a central frequency of 40 kHz.}

\added[id=CG]{To model the wave propagation for an infinite unbounded homogeneous space, we set both material 1 and material 2 to be Titanium ($V_P^{Ti}=6011.6~\text{m}/\text{s}$, $V_S^{Ti}=3093.0~\text{m}/\text{s}$, and $\rho^{Ti}=4506.0~\text{g}/\text{cm}^3$) or Granite ($V_P^{Gt}=5616.0~\text{m}/\text{s}$, $V_S^{Gt}=3463.0~\text{m}/\text{s}$, and $\rho^{Gt}=2.75~\text{g}/\text{cm}^3$), and add absorbing boundary conditions outside (orange region). The black lines in Figure \ref{fig:fd}(b) and (c) show the modeled waveforms for the baseline cases for Titanium and Granite samples with the infinite unbounded homogeneous approximation.}

\added[id=CG]{Then we set material 1 to be Steel ($V_P^{Steel}=5525.8~\text{m}/\text{s}$, $V_S^{Steel}=2927.0~\text{m}/\text{s}$, and $\rho^{Steel}=8050.0~\text{g}/\text{cm}^3$), which is the same as for the real experimental setup with the ball drop apparatus made of Steel. The received waveforms for the Titanium and Granite samples are shown in dashed red lines in Figure \ref{fig:fd}(b) and (c), compared with the relative corresponding infinite unbounded approximation. The waveforms are almost identical for the Titanium sample with or without the Steel--Titanium boundary. The waveform difference is more obvious for the Granite sample. This is because the impedance contrast between Granite and Steel is bigger than that between Titanium and Steel.}

\end{document}